\documentclass[12pt,review]{elsarticle}
\usepackage[top=0.8in, bottom=0.8in, left=0.9in, right=0.9in]{geometry}

\usepackage{subfigure}
\usepackage{tabularx}

\usepackage{amsmath}
\usepackage{amssymb}
\usepackage{amsthm}

\usepackage{lineno}
\usepackage{comment}
\usepackage[unicode]{hyperref}

\journal{Elsevier}
\begin{document}

\begin{frontmatter}



\title{Multiscale discrete Maxwell boundary condition for the discrete unified gas kinetic scheme for all Knudsen number flows}


\author[inst1]{Ziyang Xin}

\affiliation[inst1]{organization={State Key Laboratory of Coal Combustion, School of Energy and Power Engineering},%Department and Organization
            addressline={Huazhong University of Science and Technology}, 
            city={Wuhan},
            postcode={430074}, 
            %%state={State One},
            country={China}}

\author[inst2]{Yue Zhang}
\author[inst3]{Chuang Zhang}
\author[inst1,inst4]{Zhaoli Guo\corref{cor1}}

\affiliation[inst2]{organization={Hubei Provincial Engineering Technology Research Center of Green Chemical Equipment, School of Mechanical and Electrical Engineering,},%Department and Organization
            addressline={Wuhan Institute of Technology}, 
            city={Wuhan},
            postcode={430205}, 
            %%state={State One},
            country={China}}
\affiliation[inst3]{organization={Department of Physics},%Department and Organization
            addressline={Hangzhou Dianzi University}, 
            city={Hangzhou},
            postcode={310018}, 
            %%state={State One},
            country={China}}
\affiliation[inst4]{organization={Institute of Interdisciplinary Research for Mathematics and Applied Science},%Department and Organization
            addressline={Huazhong University of Science and Technology}, 
            city={Wuhan},
            postcode={430074}, 
            %%state={State One},
            country={China}}
\cortext[cor1]{Corresponding author. 
Email address: zlguo@hust.edu.cn
}
\begin{abstract}
In this paper, a multiscale boundary condition for the discrete unified gas kinetic scheme (DUGKS) is developed for gas flows in all flow regimes. Based on the discrete Maxwell boundary condition (DMBC), this study addresses the limitations of the original DMBC used in DUGKS. Specifically, it is found that the DMBC produces spurious velocity slip and temperature jump, which are proportional to the mesh size and the momentum accommodation coefficient.
The proposed multiscale DMBC is implemented by ensuring that the reflected original distribution function excludes collision effects. Theoretical analyses and numerous numerical tests show that the multiscale DMBC can achieve exactly the non-slip and non-jump conditions in the continuum limit and accurately captures non-equilibrium phenomena across a wide range of Knudsen numbers. The results demonstrate that the DUGKS with the multiscale DMBC can work properly for wall boundary conditions in all flow regimes with a fixed discretization in both space and time, without limitations on the thickness of the Knudsen layer and relaxation time.

\end{abstract}




\end{frontmatter}

%\linenumbers

\section{Introduction} \label{sec:intro}
Gas flows interacting with solid surfaces or structures are widely encountered in many applications, such as spacecraft re-entry~\cite{blanchard1997aerodynamic, schouler2020survey}, unconventional gas extraction~\cite{wang2014natural, moghaddam2016slip}, and micro-electro-mechanical systems~\cite{karniadakis2006microflows,regmi2018micro}. In these applications, the systems usually encounter a wide span of characteristic lengths or pressure conditions.
To accurately model and simulate these scenarios, it is essential to adopt proper governing equations and boundary conditions (BCs). The Knudsen number ($\text{Kn}$), defined as the ratio of the mean free path to the characteristic length of the system, is commonly employed to assess the applicability of the different governing equations and BCs. In general, the traditional Navier-Stokes-Fourier (NSF) equations, incorporating with no-slip and no-jump conditions, are valid in the continuum regime ($\text{Kn} < 0.001$). When $0.001 \leq \text{Kn} < 0.1$, the flow is in the slip regime, and the NSF equations, accompanied with appropriate velocity-slip and temperature-jump wall boundary conditions, can predict certain main features of the flow~\cite{guo2014generalized, zeng2023second}. However, for flows in transition regime ($0.1 \leq \text{Kn} < 10$) and free-molecular regime ($\text{Kn} \geq 10$), the NSF equations fail to work because in that the continuum assumption no longer holds.
To capture gas flow behaviors across multiple flow regimes, the Boltzmann equation or its simplified kinetic models can serve as a good model~\cite{chapman1990mathematical}.

In recent decades, numerous numerical methods based on the kinetic model equations were developed, such as the lattice Boltzmann method (LBM)~\cite{guo2013lattice}, direct simulation Monte Carlo (DSMC) method~\cite{bird1994molecular,fan2001statistical}, and discrete velocity method (DVM)~\cite{broadwell1964study}. The well-known LBM is mainly used to simulate continuum flows with low Mach number, which provides the benefits of easy implementation of BCs on complex geometric structures and parallel computing~\cite{guo2013lattice}. Many attempts have been made to extend the LBM to simulate gas flow in the slip regime and even transition regime, such as the use of diffuse-reflection BCs~\cite{succi2002mesoscopic,ansumali2002kinetic} and effective viscosity models~\cite{guo2006physical,kim2008slip}. However, the accuracy and effectiveness of those extended LBMs in simulating rarefied gas flows in the transition and free molecular regimes are still questionable~\cite{liu2013lattice,su2017comparative}. On the contrary, the conventional DSMC and DVM methods were demonstrated to be able to accurately solve flow problems in all regimes~\cite{bird1994molecular,mieussens2000discrete,john2011effects,meng2013assessment,akhlaghi2023comprehensive}. However, the requirement that the cell size and time step must be respectively smaller than the mean free path and relaxation time, owing to the decoupled treatment of molecular collision and streaming, renders both methods computationally expensive in the continuum or near-continuum regimes. To circumvent these limitations, some multiscale methods were developed with the asymptotic preserving (AP) properties toward the Euler limit~\cite{jin1999efficient,filbet2010class,dimarco2013asymptotic}. Such numerical methods can capture the flow behavior across multiple flow regimes with a fixed discretization in both space and time, without the limitations of the mean free path and relaxation time. 

Furthermore, the unified gas-kinetic scheme (UGKS) for all flow regimes, which is a second-order unified preserving (UP) scheme~\cite{guo2023unified} capable of accurately simulating flow behaviors at the NSF limit, was proposed~\cite{xu2010unified,huang2012unified}. It is a finite volume method that utilizes a local time-integral solution of the kinetic equation to reconstruct numerical fluxes at the interface. Based on a similar principle, the discrete unified gas-kinetic scheme (DUGKS), which utilizes the discrete characteristic solution of the kinetic equation for the reconstruction of numerical fluxes, was subsequently developed~\cite{Guo2013DiscreteUG,guo2015discrete}. Due to the coupling treatment of the transport and collision processes in the flux reconstruction, both UGKS and DUGKS exhibit good UP properties and can capture accurately the NSF solution with cell size and time step much larger than the mean free path and relaxation time~\cite{guo2023unified}. Additionally, in DUGKS, the evaluation of the numerical flux is simplified by introducing a transformed velocity distribution function (VDF) with collision term, a technique similarly employed in the LBM. This strategy makes DUGKS simpler than UGKS and relative more efficient~\cite{wang2015unified}. However, it also changes the implementation of BCs from the traditional implementation in LBM and UGKS~\cite{yang2020analysis,chen2022maxwell}.

Different BCs in DUGKS for the gas-solid interaction have already been employed successfully
to predict the behavior of gas flow ranging from continuum to free-molecular regimes~\cite{wang2015comparative,zhang2018discrete,zhu2019application,tao2021application,gu2021computational,wang2022investigation,xin2023discrete}. In the continuum regime, the bounce-back scheme~\cite{Guo2013DiscreteUG} and non-equilibrium extrapolation scheme~\cite{wu2016discrete} were introduced from LBM to DUGKS to realize the no-slip BC. However, both schemes fail to describe the gas-solid interaction in rarefied gas flows.
In the rarefied regime, the Maxwell boundary condition (MBC)~\cite{maxwell1879vii} has been employed to capture the non-equilibrium phenomena, such as velocity slip, temperature jump, and Knudsen minimum. Several researchers have observed discrete effects resulting from the discrete Maxwell boundary condition (DMBC)~\cite{guo2007discrete,mieussens2013asymptotic}, which arises from the discretization of MBC through different numerical schemes.
In LBM, the discrete effects of DMBC were first analyzed, revealing a nonphysical slip dependent on lattice spacing~\cite{guo2007discrete}. On the other hand, the analysis of the discrete effects of BC in UGKS shows the UP properties may be broken~\cite{mieussens2013asymptotic}. Specifically, when the full diffusive DMBC is implemented in the UGKS, the cell adjacent to the solid surface must be finer than the thickness of the Knudsen layer to achieve no-slip condition in the continuum regime successfully.
To address this problem, a multiscale full diffusive DMBC was proposed in UGKS that uses a local time solution to change the incident VDF~\cite{chen2015comparative}. It was shown that the multiscale DMBC works properly across different regimes with a fixed discretization in both space and time, without limitations on the thickness of the Knudsen layer and relaxation time. 

Therefore, it is natural to ask whether such discrete effects of DMBC also exist in DUGKS, and if so, a unified and multiscale DMBC applicable to flows across different regimes becomes essential for DUGKS. Due to the different evolution procedure and flux evaluation between UGKS and DUGKS, the multiscale DMBC for UGKS cannot be applied directly in DUGKS.  On the other hand, the multiscale DMBC in UGKS only addresses the discrete effects of diffuse reflection part in the DMBC, without accounting for the specular reflection part.
Consequently, a multiscale DMBC for DUGKS, which considers both diffuse scattering and specular reflection of the gas, still requires further development and analysis. 

The remaining part of this paper is organized as follows. Section~\ref{sec: numerical formulation} introduces the basics of kinetic theory and DUGKS. Section~\ref{sec: Boundary condition} outlines both the original and the developed multiscale DMBC within the DUGKS framework, complete with a rigorous theoretical analysis of their discrete effects. Numerical tests of the DMBCs are conducted in Sec.~\ref{numerical tests}. Finally, the conclusions are drawn in Sec.~\ref{conclusion}

\section{Numerical Formulation} \label{sec: numerical formulation}
\subsection{Relaxation model of the Boltzmann equation }
In this section, some basics of kinetic theory and MBC for monatomic gases are introduced. To avoid the mathematical difficulty caused by the original nonlinear integral collision term, the Boltzmann model equation with the relaxation collision term is generally adopted, 
\begin{equation} \label{eq:BGK}
    \frac { \partial f } { \partial t } + {\boldsymbol \xi} \cdot  \nabla f = \Omega ( f ) \equiv \frac { g - f } { \tau } , 
\end{equation}
where $ f\equiv f( {\boldsymbol x},{\boldsymbol \xi}, t)$ is the velocity distribution function (VDF) with particle velocity $ {\boldsymbol \xi} \equiv ( \xi_x, \xi_y, \xi_z)$ at position ${\boldsymbol x}\equiv( x, y, z)$ and time $ t$. $\Omega( f)$ is the collision operator, $g$ is the equilibrium VDF and $ \tau$ is the relaxation time relating to the dynamic viscosity $\mu$ and pressure $p$ with $\tau = \mu/p$. For monatomic gases, the Bhatnagar-Gross-Krook (BGK)~\cite{bhatnagar1954model}, Ellipsoidal Statistical (ES)~\cite{holway1966new}, and Shakhov-BGK~\cite{shakhov1968generalization} models can be formulated in the above relaxation form. 

The conservative flow variables $\boldsymbol W$, including the density $\rho$, flow velocity $\boldsymbol{u}$, and total energy $\rho E$, are defined by the moments of the VDF, 
\begin{equation}\label{eq: macro variable}
    \boldsymbol W = \begin{pmatrix} \rho \\ \rho{\boldsymbol u} \\ \rho E \end{pmatrix} = \int \boldsymbol \psi({\boldsymbol \xi}) f d{\boldsymbol \xi} =\left \langle \boldsymbol \psi({\boldsymbol \xi}), f \right \rangle ,
\end{equation}
where  $ \rho E = \rho ( u^2 + 3R T)/2 $ with $R$ being the gas constant, $T$ is the temperature, and $\boldsymbol \psi (\boldsymbol \xi) = \left(1,{\boldsymbol \xi}, { \xi}^2/2\right)^T$ is the collision invariant. The pressure is calculated by the ideal-gas equation of state, $p=\rho R T$.
The heat flux vector and stress tensor are defined by,
\begin{equation} \label{eq:high order macro}
    {\boldsymbol q}  = \left \langle \frac{1}{2}  \boldsymbol{c} \boldsymbol{c}^2, f  \right \rangle , \quad \quad
   {\boldsymbol \sigma} = \left \langle \boldsymbol{c} \boldsymbol{c}, f- f^{eq}  \right \rangle,
\end{equation}
where $\boldsymbol{c} = \boldsymbol{\xi} - \boldsymbol{u}$ is the peculiar velocity.

The viscosity $\mu$ can be assumed to follow a power law dependence on temperature, 
\begin{equation}
    \mu = \mu_0 \left(\frac{T}{T_0}\right)^\theta,
\end{equation}
where $\theta$ is related to the molecular interaction model, $\mu_0$ is the reference viscosity at the reference temperature $T_0$, and the reference mean free path can be calculated via $\lambda_0 = \tau_0 \sqrt{\pi RT_0/2}$. The Knudsen number is defined as follows,
\begin{equation}
    \text{Kn} =\frac{\lambda_0}{L} =\frac{\mu_0 }{p_0 L}\sqrt{\frac{\pi RT_0}{2}},
\end{equation}
where $L$ is the characteristic length of the system.

Additionally, the MBC is used in association with the kinetic equation~\eqref{eq:BGK} to describe gas-solid interaction. In particular, when gas molecules are incident on solid surface, a portion of molecules ($\alpha$) undergo diffuse reflection while the remainder $\left(1-\alpha\right)$ will be reflected specularly. Here, $\alpha$ is known as the tangential momentum accommodation coefficient of the wall. The cases of $\alpha = 1$ and $\alpha = 0$ correspond to the fully diffuse-reflection and specular-reflection conditions, respectively. Then the VDF of reflected gas molecules can be expressed as
\begin{equation}\label{eq:MBC}
    f_{re} = \alpha f^{eq}_w + \left(1-\alpha \right)f_{in}\left( \boldsymbol \xi_{t}, -\xi_{n}\right),\quad \quad \xi_n > 0,
\end{equation}
where $f_{in}$ is the VDF of incident gas molecules. The subscripts $n$ and $t$ denote the normal and tangential velocity components, respectively. Particularly, $\xi_{n}=\boldsymbol \xi\cdot \boldsymbol n_w$ and $\boldsymbol \xi_{t}=\boldsymbol \xi- \xi_{n}\boldsymbol n_w$ represent the normal and tangential component of velocity $\boldsymbol \xi$, respectively, with $\boldsymbol n_w$ being the unit vector normal to the wall pointing to the cell. Assuming the wall moves with a tangential velocity $\boldsymbol u_w$ and has a temperature $T_w$, the half-range Maxwellian equilibrium is
\begin{equation}
    f^{eq}_w=\frac{\rho_w}{(2\pi R T_w)^{3/2}}\exp \left(-\frac{|\boldsymbol \xi - \boldsymbol u_w|^ {2}}{2 R T_w}\right),\quad \quad \xi_n > 0,
\end{equation}
with $\rho_w$ representing the reflected density at the wall, which can be calculated from the non-penetration condition (i.e., the mass flow normal to the wall is equal to zero).  In this study, the normal component of $\boldsymbol u_w$ is set to 0, i.e., $ u_{w,n}=0$. 

\subsection{Discrete unified gas-kinetic scheme}\label{subsec: BGK model}
The DUGKS is a cell-centered finite volume method to discretize the kinetic equation~\eqref{eq:BGK}. Initially, the physical space is divided into a set of cells $V_j$ centered at $\boldsymbol x_j$. Then integrating Eq.~\eqref{eq:BGK} over cell $V_j$ from time $t^n$ to time $t^{n+1}=t^{n}+\Delta t$, and using the trapezoidal and middle-point rules for the time integration of the collision and convection terms, respectively, we can obtain the following relation: 
\begin{equation}\label{eq:dugks equation}
    \tilde{f}^{n+1}_j= \tilde{f}^{+,n}_j  -\frac{\Delta t}{|V_j|} \boldsymbol F^{n+1/2},
\end{equation}
with 
\begin{equation} \label{eq:transform}
    \tilde{f}=f-\frac{\Delta t}{2} \Omega, \quad
    \tilde{f}^+=f+\frac{\Delta t}{2} \Omega,
\end{equation}
are two new distribution functions to remove the implicity.  The microscopic flux across the cell interface at the half time $t^{n+1/2} = t^n + \Delta t/2$ is evaluated by the Gauss theorem,
\begin{equation} \label{interface flux}
    \boldsymbol F^{n+1/2} (\boldsymbol \xi) = \int_{\partial V_j}(\boldsymbol \xi \cdot \boldsymbol n) f^{n+1/2} d \boldsymbol S,
\end{equation}
where $\partial V_j$ is the surface of cell $V_j$, and $\boldsymbol n$ is the outward unit vector normal to the interface. 
%Note that not discretizing the velocity space at this stage considerably simplifies the notation

Due to the conservation properties of the collision term, the conservative variables $\boldsymbol W$ can also be calculated from
\begin{equation}\label{eq:W tidle}
    \boldsymbol W  = \left \langle \boldsymbol \psi, \tilde{f}  \right \rangle.
\end{equation}
In addition, $\tilde{f}_j^{+,n}$ at time $t^n$ can be obtained from $\tilde{f}_j^{n}$ according to Eq.~\eqref{eq:transform},
\begin{equation}\label{eq:tidle f n}
    \tilde{f}_j^{+,n} = \frac{2\tau - \Delta t}{2\tau + \Delta t}\tilde{f}_j^{n} + \frac{2\Delta t}{2\tau + \Delta t} g^{n}_j,
\end{equation}
where the equilibrium VDF $g^{n}$ is evaluated using $\boldsymbol W^n$, which is also calculated from $\tilde{f}_j^{n}$. Therefore, we can track the evolution of $\tilde{f}$ instead of the original VDF in the implementation, which can evolve explicitly according to Eq.~\eqref{eq:dugks equation} when $f^{n+1/2}$ is known.

\begin{figure}[htbp]
\centering
\includegraphics[width=0.6\textwidth]{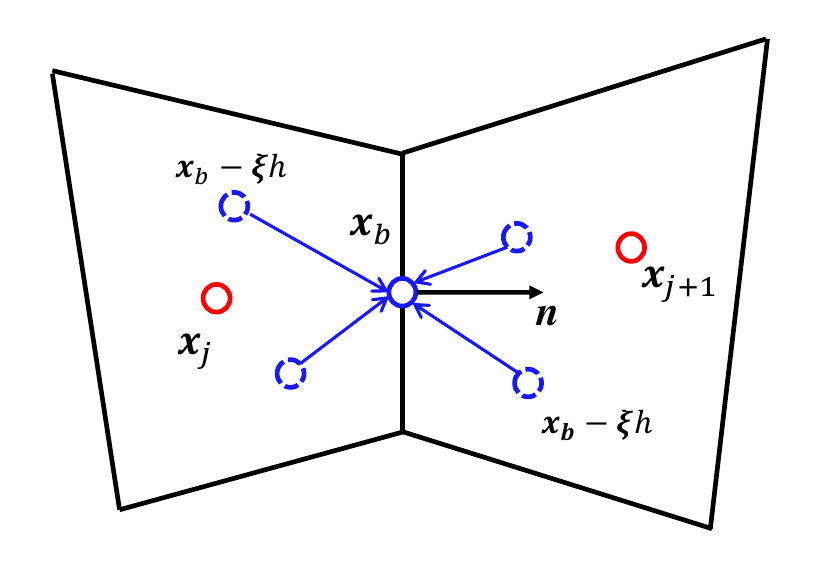}
\caption{
 Schematic of two neighboring cells $V_j$ and $V_{j+1}$ centered at $\boldsymbol x_j$ and $\boldsymbol x_{j+1}$, respectively. $\boldsymbol x_b $ is the midpoint of the cell interface, and $\boldsymbol n$ is the outward norm unit vector of cell $V_{j}$ at $\boldsymbol x_b $.
}\label{FIG_DUGKS_characteristicLine}
\end{figure}

The key procedure of DUGKS is use of the discrete characteristic solution of the kinetic equation to reconstruct the VDF at the interface. Specifically, the kinetic equation~\eqref{eq:BGK} is integrated along the characteristic line  in a half-time step $h=\Delta t/2$ from $\boldsymbol x_b - \boldsymbol \xi h$ to $\boldsymbol x_b$, and employing the trapezoidal rule to evaluate the collision term, we can get 
\begin{equation} \label{eq:characteristic line}
    f^{n+1/2}(\boldsymbol x_b) - f^n(\boldsymbol x_b - \boldsymbol \xi h) = \frac{h}{2}\left[\Omega^{n+1/2}(\boldsymbol x_b) + \Omega^n(\boldsymbol x_b - \boldsymbol \xi h) \right],
\end{equation}
where $\boldsymbol x_b$ is the midpoint of the cell interface as shown in Fig.~\ref{FIG_DUGKS_characteristicLine}. Again the implicitness can be avoided by introducing another two transformed VDF,
\begin{equation} \label{eq:transformbar}
    \bar{f}=f-\frac{h}{2} \Omega, \quad
    \bar{f}^+=f+\frac{h}{2} \Omega,
\end{equation}
and then Eq.~\eqref{eq:characteristic line} can be rewritten as
\begin{equation} \label{eq:transform characteristic line}
    \bar{f}^{n+1/2}(\boldsymbol x_b) = \bar{f}^{+,n}(\boldsymbol x_b - \boldsymbol \xi h).
\end{equation}

Here, the transformed VDF $\bar{f}^{+,n}(\boldsymbol x_b - \boldsymbol \xi h)$ can be interpolated linearly from the cell-averaged distribution functions $\bar{f}^{+,n}$ at its neighboring cell centers and then substituted into Eq.~\eqref{eq:transform characteristic line}, resulting in
\begin{equation}\label{eq:fbar}
\begin{aligned}
    \bar{f}^{n+1/2}(\boldsymbol x_b) = \begin{cases}
    \bar{f}^{+,n}(\boldsymbol x_j) + (\boldsymbol x_b - \boldsymbol \xi h -\boldsymbol x_j) \cdot \nabla \bar{f}^{+,n}(\boldsymbol x_j), &\boldsymbol \xi \cdot  \boldsymbol n > 0, \\
    \bar{f}^{+,n}(\boldsymbol x_{j+1}) + (\boldsymbol x_b - \boldsymbol \xi h -\boldsymbol x_{j+1}) \cdot \nabla \bar{f}^{+,n}(\boldsymbol x_{j+1}), & \boldsymbol \xi \cdot  \boldsymbol n < 0,
    \end{cases}
\end{aligned}
\end{equation}
where $\bar{f}^{+,n}$ is calculated from the tracked $\tilde{f}^{n}$, 
\begin{equation} \label{eq:fbar+}
    \bar{f}^{+,n} = \frac{2\tau - h}{2\tau + \Delta t}\tilde{f}^{n} + \frac{3h}{2\tau + \Delta t}g^{n},
\end{equation}
and the slope $\nabla \bar{f}^{+,n}$ is calculated from the value of cell-averaged $\bar{f}^{+,n}$  by employing a central difference scheme for continuous flows or van Leer limiter for flows with discontinuities~\cite{Guo2013DiscreteUG,guo2015discrete}. The macroscopic variables at the cell interface are obtained from
\begin{equation}\label{W n+1/2}
     \boldsymbol W^{n+1/2}(\boldsymbol x_b)  = \left \langle \boldsymbol \psi, \bar{f}^{n+1/2}(\boldsymbol x_b)  \right \rangle,
\end{equation}
which is used in $g^{n+1/2}$ to obtain the original VDF at the cell interface according to the
relationship in Eq.~\eqref{eq:transformbar}, 
\begin{equation} \label{eq:updating rule of Original f}
    f^{n+1/2}(\boldsymbol x_b) = \frac{2\tau}{2\tau + h}\bar{f}^{n+1/2}(\boldsymbol x_b) + \frac{h}{2\tau + h}g^{n+1/2}(\boldsymbol x_b). 
\end{equation}
Then, the interface flux can be determined based on Eq.~\eqref{interface flux}.

\section{Boundary conditions}\label{sec: Boundary condition}
Section~\ref{sec: numerical formulation} presents the numerical formulation of DUGKS for interior cells, based on kinetic equation~\eqref{eq:BGK}. The distinctive aspect of DUGKS is the evolution of the transformed VDF $\bar f$ at cell interfaces as outlined in Eq.~\eqref{eq:fbar}. As a result, the original DMBC is a discretization of the MBC~\eqref{eq:MBC} in DUGKS to determine the unknown reflected $\bar f$ at the wall. Motivated by previous studies on the discrete effects of DMBC within LBM and UGKS~\cite{guo2007discrete,mieussens2013asymptotic,chen2015comparative}, we now explore the discrete effects of the original DMBC employed in DUGKS and propose a multiscale DMBC for all Knudsen number flows.

\subsection{Original discrete Maxwell boundary condition}\label{KBC}
The original DMBC in DUGKS is implemented to determine the unknown transformed VDF $\bar f$ at a wall point $\boldsymbol x_w$ at time $t^{n+1/2}$,
\begin{equation}\label{eq:method I fbar}
\begin{aligned}
    \bar{f}^{n+1/2}(\boldsymbol x_w) = \begin{cases}
    \bar{f}_{in}^{n+1/2}(\boldsymbol x_w, \boldsymbol \xi_{t}, \xi_{n}), &\xi_n< 0, \\
    \alpha f^{eq}_w + \left(1-\alpha \right)\bar{f}_{in}^{n+1/2}\left(\boldsymbol x_w, \boldsymbol \xi_{t}, -\xi_{n}\right), & \xi_n > 0,
    \end{cases}
\end{aligned}
\end{equation}
where $\bar{f}_{in}^{n+1/2}(\boldsymbol x_w)$ represents the transformed VDF of the incident molecules, which is calculated by Eq.~\eqref{eq:fbar}. $\bar{f}_{in}$ is regarded known prior to the implementation of BC unless otherwise stated. 
Thus, $\rho_w$ is the only unknown quantity in $f^{eq}_w$ as given by Eq.~\eqref{eq:method I fbar}. Here, the condition of zero mass flux normal to the wall is employed to calculate the reflected density,
\begin{equation}
\left \langle \xi_n,\bar{f}_{in}^{n+1/2}(\boldsymbol x_w, \boldsymbol \xi_{t}, \xi_{n}) \right \rangle _{<0} + \left \langle \xi_n, \alpha f^{eq}_w + \left(1-\alpha \right)\bar{f}_{in}^{n+1/2}\left(\boldsymbol x_w, \boldsymbol \xi_{t}, -\xi_{n}\right) \right \rangle _{>0} = 0,
\end{equation}
which gives
\begin{equation} \label{rho_w method I}
\rho_w = -\sqrt{\frac{2\pi}{RT_w}}\left \langle \xi_n, \bar{f}_{in}^{n+1/2}(\boldsymbol x_w, \boldsymbol \xi_{t}, \xi_{n}) \right \rangle _{<0},
\end{equation}
where the symbol $\left \langle \cdot ,  \cdot\right \rangle _{<0} $ represents the integral for the  normal component of the velocity $\xi_n$ in the range $(-\infty, 0)$ and the $\left \langle \cdot ,  \cdot \right \rangle _{>0} $ represents the integral for the  normal component of the velocity $\xi_n$ in the range $(0,\infty)$. Then, the transformed VDF $\bar f^{n+1/2}(\boldsymbol x_w)$ is completely determined and can be used to calculate the macroscopic flow variables and the original VDF at the interface $\boldsymbol x_w$ at time $t^{n+1/2}$.

According to Eq.\eqref{W n+1/2}, the macroscopic flow variables can be written as,
\begin{equation} \label{W method I}
        \boldsymbol{W}^{n+1/2}(\boldsymbol{x}_w) = \left \langle \boldsymbol \psi, \bar{f}_{in}^{n+1/2} (\boldsymbol \xi_{t}, \xi_{n})  \right \rangle_{<0} + \left(1-\alpha\right) \left \langle \boldsymbol \psi, \bar{f}_{in}^{n+1/2} (\boldsymbol \xi_{t}, -\xi_{n})  \right \rangle_{>0} + \frac{\alpha}{2} \boldsymbol{I}_w,
\end{equation}
where the vectors are rewritten as,
\begin{equation}
\begin{aligned}
    \boldsymbol W &= \left(\rho, \rho u_n, \rho{\boldsymbol u}_t, \rho E\right)^T, \quad  \quad \quad \quad
     \boldsymbol{\psi} = \left( 1, \xi_n, \boldsymbol \xi_t, \frac{1}{2} \xi ^2 \right)^T,\\
    \boldsymbol I_w &=  \left( \rho_w, \rho_w  \sqrt{\frac{2RT_w}{\pi}}, \rho_w{\boldsymbol u}_{w,t}, \frac{3}{2} \rho_w RT_w+ \frac{1}{2}\rho_wu_w^2 \right)^T.
\end{aligned}
\end{equation}
Note that $u_n = 0$ due to the non-penetration condition.
% It can be seen that Method I is easy to implement hence it is widely used in the DUGKS~\cite{tao2021application,tao2020ghost}. 

In the end, the original VDF at the interface $\boldsymbol x_w$ can be obtained by
\begin{equation}\label{eq:f x_w Method I}
\begin{aligned}
    f^{n+1/2}(\boldsymbol x_w) = \begin{cases}
    \frac{2\tau}{2\tau + h}\bar{f}_{in}^{n+1/2}(\boldsymbol x_w,\boldsymbol \xi_{t}, \xi_{n}) + \frac{h}{2\tau + h}g^{n+1/2}(\boldsymbol x_w), &\xi_n < 0, \\
   \frac{2\tau}{2\tau + h}\left(  \alpha f^{eq}_w+\left( 1- \alpha\right)\bar{f}_{in}^{n+1/2}(\boldsymbol x_w,\boldsymbol \xi_{t}, -\xi_{n}) \right)+\frac{h}{2\tau + h}g^{n+1/2}(\boldsymbol x_w), & \xi_n > 0,
    \end{cases} 
\end{aligned}
\end{equation}
where $g^{n+1/2}$ depends on the macroscopic flow variables $\boldsymbol{W}^{n+1/2}(\boldsymbol{x}_w)$ of gas at the wall, rather than the density $\rho_w$, velocity $\boldsymbol{u}_w$ and temperature $T_w$ of the wall. The original DMBC in DUGKS is widely used for rarefied flow simulations due to its simple implementation~\cite{tao2021application,tao2020ghost}. However, it violates the no-slip and no-jump conditions in the continuum limit due to the collision effect being coupled to the reflected VDF, as analyzed in Sec.~\ref{Continuum limit}.

\subsection{Multiscale discrete Maxwell boundary condition} \label{MBC}
To overcome the limitations of the original DMBC, we propose a multiscale DMBC for DUGKS that functions effectively across all flow regimes. Specifically, the multiscale DMBC is utilized directly to determine the unknown original VDF in Eq.~\eqref{eq:updating rule of Original f} at the cell interface $\boldsymbol x_w$ at time $t^{n+1/2}$, 
\begin{equation}\label{eq:f x_w}
\begin{aligned}
    f^{n+1/2}(\boldsymbol x_w) = \begin{cases}
    \frac{2\tau}{2\tau + h}\bar{f}_{in}^{n+1/2}(\boldsymbol x_w,\boldsymbol \xi_{t}, \xi_{n}) + \frac{h}{2\tau + h}g^{n+1/2}(\boldsymbol x_w), &\xi_n < 0, \\
    \alpha f^{eq}_w+\left( 1- \alpha\right)f_{in}^{n+1/2}(\boldsymbol x_w,\boldsymbol \xi_{t}, -\xi_{n}) , & \xi_n > 0,
    \end{cases} 
\end{aligned}
\end{equation}
with
\begin{equation} \label{Multiscale BC f}
    f_{in}^{n+1/2}(\boldsymbol x_w,\boldsymbol \xi_{t}, -\xi_{n}) = \frac{2\tau}{2\tau + h}\bar{f}_{in}^{n+1/2}(\boldsymbol x_w,\boldsymbol \xi_{t}, -\xi_{n}) + \frac{h}{2\tau + h}g^{n+1/2}(\boldsymbol x_w), \quad \xi_n >0,
\end{equation}
where the unknowns $\rho_w$ and $\boldsymbol W^{n+1/2}(\boldsymbol x_w)$ are used in the equilibrium distribution function $f^{eq}_w$ and $g^{n+1/2}$ as given by the above equations. First, the condition of zero mass flux normal to the wall is still applied to ensure that the wall is impenetrable,
\begin{equation}\label{eq: f zero mass flux}
\left \langle \xi_n,f^{n+1/2}(\boldsymbol x_w,\boldsymbol \xi_{t}, \xi_{n}) \right \rangle _{<0} + \left \langle \xi_n ,\alpha f^{eq}_w+\left( 1- \alpha\right)f_{in}^{n+1/2}(\boldsymbol x_w,\boldsymbol \xi_{t}, -\xi_{n})\right \rangle _{>0} = 0.
\end{equation}
Further, the corresponding moment relations are introduced,
\begin{equation}
    \boldsymbol W^{n+1/2}(\boldsymbol x_w) = \left \langle \boldsymbol \psi, f^{n+1/2}(\boldsymbol x_w) \right \rangle.
\end{equation}

With the above relations and Eqs.~\eqref{eq:f x_w} and~\eqref{eq: f zero mass flux}, we can get
\begin{equation}\label{rho_w method II}
    \rho_w = -\sqrt{\frac{2\pi}{RT_w}}\left( \frac{2\tau}{2\tau+h}\left \langle \xi_n,\bar{f}_{in}^{n+1/2}(\boldsymbol \xi_{t}, \xi_{n}) \right \rangle _{<0} +\frac{h}{2\tau+h} \left \langle \xi_n, g^{n+1/2}(\boldsymbol x_w) \right  \rangle_{<0} \right),
\end{equation}
and
\begin{equation} \label{eq: nonlinear W}
\begin{aligned}
    \boldsymbol{W}^{n+1/2}(\boldsymbol{x}_w) =& \frac{2\tau}{2\tau+h} \left \langle \boldsymbol \psi ,\bar{f}_{in}^{n+1/2}(\boldsymbol \xi_{t}, \xi_{n})  \right \rangle_{<0} +  \frac{h}{2\tau+h} \left \langle \boldsymbol \psi, g^{n+1/2}(\boldsymbol x_w) \right  \rangle_{<0} 
    \\& + (1-\alpha)\left \langle \boldsymbol \psi ,f_{in}^{n+1/2}(\boldsymbol \xi_{t}, -\xi_{n})  \right \rangle_{>0} + \frac{\alpha}{2} \boldsymbol{I}_w .
\end{aligned}
\end{equation}
The above system of equations with respect to the variables $\rho_w$ and $\boldsymbol W^{n+1/2}(\boldsymbol x_w)$ is nonlinear, which can be solved using certain iteration methods (the nonlinear Gauss-Seidel method is employed here). However, for isothermal case (i.e., $T=T_w = T_0$), $\rho_w$ and $\boldsymbol W^{n+1/2}$can be solved analytically 
\begin{equation} 
\begin{aligned}
      \rho_w &= \left(2 - \alpha\right)\frac{h}{2\tau + h}\left \langle 1,\bar{f}_{in}^{n+1/2}\right \rangle_{<0}  +\frac{4\tau+\alpha h}{2\tau + h} \sqrt{\frac{\pi}{2RT_0}}\left |\left \langle \xi_n,\bar{f}_{in}^{n+1/2}\right \rangle_{<0} \right|,\\
      \rho^{n+1/2} &= (2-\alpha)\left \langle 1,\bar{f}_{in}^{n+1/2}\right \rangle_{<0}+\alpha\sqrt{\frac{\pi}{2RT_0}}\left |\left \langle \xi_n,\bar{f}_{in}^{n+1/2}\right \rangle_{<0} \right|,\\
      u^{n+1/2}_n &= 0,\\
      \rho^{n+1/2}\boldsymbol{u}^{n+1/2}_t &=\frac{4\tau(2-\alpha)}{4\tau + \alpha h}\left \langle \boldsymbol{\xi}_t,\bar{f}_{in}^{n+1/2}\right \rangle_{<0} + \frac{ \alpha  \left(2\tau + h\right)}{4\tau + \alpha h} \rho_w \boldsymbol{u}_{w,t}.
\end{aligned}
\end{equation}
In the end, the interface flux at the wall $\boldsymbol x_w$ can be determined based on Eq.~\eqref{eq:f x_w}. 

In this work, the multiscale DMBC is implemented in DUGKS by ensuring that the reflected original VDF excludes collision effects. This approach contrasts with the multiscale DMBC used in UGKS, where a local time solution modifies the incident VDF~\cite{chen2015comparative}.

\subsection{Analysis of original and multiscale DMBCs}\label{BC analysis}
The differences between the original and multiscale DMBCs for the distribution function at the wall will be analyzed in this section. The gas transport in the free-molecular and continuum limits are particularly investigated. Without loss of generality, the BGK model is used as an example, i.e.,
\begin{equation}\label{eq: feq}
    g=f ^ {eq} = \frac { \rho } { ( 2 \pi R T ) ^ { 3 / 2 } } \exp \left( - \frac { |{\boldsymbol \xi} -{\boldsymbol u} | ^ { 2 } } { 2 R T } \right),
\end{equation}
in the continuum limit, the BGK model can recover to the NSF equations through the Chapman-Enskog (CE) expansion~\cite{chen2015comparative, kremer2010introduction}. In the process of recovering the NSF equations, the VDF $f$ is approximated by the first-order CE expansion,
\begin{equation}\label{eq:1-order CE}
    f= g-\tau D g,
\end{equation}
where $D\equiv\partial_t + \boldsymbol \xi \cdot \nabla$, and $Dg$ can be written as
\begin{equation} \label{eq:CE}
\begin{aligned}
        D g
        =f^{eq} \Bigl[ \left( \frac{c^2}{2RT} - \frac{5}{2}\right)  \frac{\boldsymbol c \cdot \nabla T}{T}
        + \frac{\boldsymbol c\cdot (\nabla \boldsymbol u) \cdot \boldsymbol c}{RT} -\frac{c^2}{3RT}\nabla \cdot \boldsymbol u\Bigr].
\end{aligned}
\end{equation}

\subsubsection{Free-molecular limit}
In the free-molecular limit as $\tau \gg \Delta t$, identical results are derived from both the original DMBC (Eqs.~\eqref{eq:method I fbar} and ~\eqref{rho_w method I}) and the multiscale DMBC (Eqs.~\eqref{eq:f x_w} and ~\eqref{rho_w method II}),
\begin{equation}
\begin{aligned}
    f^{n+1/2}(\boldsymbol x_w) = \begin{cases}
    \bar{f}_{in}^{n+1/2}(\boldsymbol x_w,\boldsymbol \xi_{t}, \xi_{n}) , &\xi_n < 0, \\
    \alpha f^{eq}_w+\left( 1- \alpha\right)\bar{f}_{in}^{n+1/2}(\boldsymbol x_w,\boldsymbol \xi_{t}, -\xi_{n}) , & \xi_n > 0,
    \end{cases}
\end{aligned}
\end{equation}
with 
\begin{equation}
\rho_w = -\sqrt{\frac{2\pi}{RT_w}}\left \langle \xi_n, \bar{f}_{in}^{n+1/2}(\boldsymbol x_w, \boldsymbol \xi_{t}, \xi_{n}) \right \rangle _{<0}.
\end{equation}
Given that $f^{n+1/2}(\boldsymbol x_w)$ is determined to be identical, the macroscopic quantities $\boldsymbol W^{n+1/2}(\boldsymbol x_w)$ are also equivalent, and their expressions are consistent with Eq.~\eqref{W method I}. As a result, both DMBCs can be regarded as equivalent in the free-molecular limit as $\tau \gg \Delta t$.

\subsubsection{Continuum limit} \label{Continuum limit}
In the continuum limit as $\tau \ll \Delta t$, the VDF at the wall interface given by Eq.~\eqref{eq:f x_w Method I} in the original DMBC  goes to 
\begin{equation}\label{eq:original f tau=0 I}
\begin{aligned}
    f^{n+1/2}(\boldsymbol x_w) = \begin{cases}
    g^{n+1/2}(\boldsymbol x_w), &\xi_n < 0, \\
    g^{n+1/2}(\boldsymbol x_w), & \xi_n > 0,
    \end{cases} 
\end{aligned}
\end{equation}
where the equilibrium state $g^{n+1/2}(\boldsymbol x_w)$ is evaluated using $\boldsymbol W^{n+1/2}$ calculated by Eq.~\eqref{W method I}. Furthermore, the first-order solution of the CE expansion of $\bar{f}_{in}^{n+1/2}(\boldsymbol x_w)$ in Eq.~\eqref{W method I} is given, which is used to determine whether the original DMBC satisfies the conventional no-slip and no-jump in the continuum limit. From Eqs.~\eqref{eq:transformbar} and~\eqref{eq:CE}, the first-order solution of CE expansion of $\bar{f}_{in}^{n+1/2}(\boldsymbol x_w)$ is expressed as
\begin{equation} \label{eq:fbar CE}
    \bar{f}_{in}^{n+1/2}(\boldsymbol x_w)=g^{n+1/2}(\boldsymbol x_w)-\frac{2\tau+h}{2}Dg^{n+1/2}(\boldsymbol x_w). 
\end{equation}

Substituting Eq.~\eqref{eq:fbar CE} into Eq.~\eqref{rho_w method I} and ~\eqref{W method I}, we have
\begin{equation} \label{CE rho_w I}
    \rho_w=\rho^{n+1/2}\sqrt{\frac{T^{n+1/2}}{T_w}}+\left(2\tau + h \right){\sqrt{\frac{\pi}{2RT_w}}} 
    \left | \left \langle \xi_n, Dg^{n+1/2}(\boldsymbol x_w) \right \rangle _{<0} \right|,
\end{equation}
and
\begin{equation}\label{CE W I}
\boldsymbol{W}^{n+1/2}(\boldsymbol x_w) = \boldsymbol{I}_w -\frac{2 - \alpha}{\alpha} \left(2\tau + h \right)\left \langle \boldsymbol \psi, Dg^{n+1/2}(\boldsymbol x_w)\right \rangle_{<0}.
\end{equation}
As $\tau \to 0$, the gas velocity and temperature at the wall $\boldsymbol x_w$ are given as,
\begin{equation}\label{spurious W I}
\begin{aligned}
    \rho^{n+1/2}&=\rho_w - \frac{2 - \alpha}{\alpha} h \left \langle 1, Dg^{n+1/2}(\boldsymbol x_w) \right \rangle_{<0},\\
    u_{n}^{n+1/2}&=0,\\
    \rho^{n+1/2}\boldsymbol u_t^{n+1/2} &=\rho_w\boldsymbol u_{w,t} - \frac{2 - \alpha}{\alpha} h \left \langle \boldsymbol \xi_t, Dg^{n+1/2}(\boldsymbol x_w) \right \rangle_{<0},\\
     \frac{3}{2}\rho^{n+1/2}RT^{n+1/2} &=\frac{1}{2}\rho_w (3RT_{w} +u_{w}^2) - \frac{1}{2}\rho^{n+1/2} (u^{n+1/2})^2  - \frac{2 - \alpha}{\alpha} h \left \langle \frac{1}{2}\xi^2, Dg^{n+1/2}(\boldsymbol x_w) \right \rangle_{<0},  
\end{aligned}
\end{equation}
it is evident that the original DMBC fails to achieve the no-slip and no-jump conditions when $\tau \to 0$ and $\alpha \neq 0 $ unless $h$ is set to be significantly smaller than $\tau$. In other words, the original DMBC in DUGKS produces a spurious velocity slip and temperature jump at the wall, both of which are of the order of magnitude $(2-\alpha)/\alpha \cdot O(h)$.

With the same analysis, the multiscale DMBC gives that
\begin{equation}\label{eq:original f tau=0}
\begin{aligned}
    f^{n+1/2}(\boldsymbol x_w) = \begin{cases}
    g^{n+1/2}(\boldsymbol x_w), &\xi_n < 0, \\
    \alpha f^{eq}_w+\left( 1- \alpha\right)f_{in}^{n+1/2}(\boldsymbol x_w,\boldsymbol \xi_{t}, -\xi_{n}) , & \xi_n > 0,
    \end{cases} 
\end{aligned}
\end{equation}
the difference between the two DMBCs becomes apparent when examining the behavior of the reflected VDF in the continuum limit. For the multiscale DMBC, the reflected VDF as specified in Eq.~\eqref{eq:original f tau=0} aligns with the MBC of kinetic theory. In contrast, the reflected VDF for the original DMBC given by Eq.~\eqref{eq:original f tau=0 I} represents an equilibrium VDF resulting from sufficient collisions with the gas molecules at the wall.

Applying the first-order solution of CE expansion of $\bar{f}_{in}$ to Eq.~\eqref{eq: nonlinear W}, we can get
\begin{equation}\label{CE rho_w II}
    \rho_w=\rho^{n+1/2}\sqrt{\frac{T^{n+1/2}}{T_w}}+2\tau{\sqrt{\frac{\pi}{2RT_w}}}   \left | \left \langle \xi_n, Dg^{n+1/2}(\boldsymbol x_w) \right \rangle _{<0} \right|,
\end{equation}
and
\begin{equation}\label{CE W II}
     \boldsymbol{W}^{n+1/2}(\boldsymbol x_w) = \boldsymbol{I}_w - \frac{2\tau(2-\alpha)}{\alpha}\left \langle \boldsymbol \psi, Dg^{n+1/2}(\boldsymbol x_w) \right \rangle_{<0}.
\end{equation}
As $\tau \to 0$ and $\alpha \neq 0 $, the gas velocity and temperature at the wall are given as,
\begin{equation}\label{spurious W II}
\begin{aligned}
    \rho^{n+1/2}(\boldsymbol x_w) &= \rho_w,\\
        u_{n}^{n+1/2}(\boldsymbol x_w)&=0,\\
    \boldsymbol u_t^{n+1/2}(\boldsymbol x_w)&=\boldsymbol u_{w,t},\\
     T^{n+1/2}(\boldsymbol x_w)&=T_{w},  
\end{aligned}
\end{equation}
it is evident that the multiscale DMBC satisfies the no-slip and no-jump conditions.

In summary, as $\Delta t \ll \tau$, both DMBCs can be regarded as equivalent. However, the original DMBC will produce a spurious velocity slip and temperature jump in the continuum limit ($\tau \ll \Delta t$), both of which are of the order of magnitude $(2-\alpha)/\alpha \cdot O(h)$. These are non-physical effects that can influence the accuracy of simulations. In contrast, the multiscale DMBC can work efficiently in the continuum limit, remaining functional without constraints from the relaxation time.
%Furthermore, unlike the original DMBC, the multiscale version ensures that the time step is not strictly limited by the relaxation time, thus offering greater flexibility in simulation settings and enhancing computational efficiency in various fluid dynamics applications.

\section{Numerical tests}\label{numerical tests}
In this section, we will assess the accuracy of the original and multiscale DMBCs in DUGKS for all flow regimes through some numerical tests. Two classical quasi-one-dimensional flow problems, i.e., the isothermal Couette flow and Fourier flow, are performed to assess the accuracy of both DMBCs in velocity slip and temperature jump. And the two-dimension lid-driven cavity flow, is further used to verify
theoretical and numerical results on the slip velocity. For isothermal cases, the equilibrium state $g$ is set to be the Maxwellian distribution function, i.e., $g=f^{eq}$; for non-isothermal case,  the equilibrium state $g$ is given by the Shakhov-BGK model, 
\begin{equation}
    g=f^S = f^{eq}\left [ 1+(1-\text{Pr})\frac{{ \boldsymbol{c} \cdot \boldsymbol q}}{5pRT} \left( \frac{\boldsymbol{c}^2}{RT}-5\right)\right],
\end{equation}
where $\text{Pr} = 2/3$ is used for monatomic gases, and $\theta$ is set to be 0.5 in hard-sphere model. The detailed process of DUGKS based Shakhov-BGK model with the original DMBC can be seen in Ref.~\cite{guo2015discrete}, and the multiscale DMBC based Shakhov-BGK model is deduced in~\ref{appendix:MBC based on Shakhov-BGK model in DUGKS}. 

The derivation of the above numerical formulation is performed in the continuous velocity space. However, for practical simulation, the velocity space is discretized into a set of discrete velocities $ \{\boldsymbol \xi_i|i=1,2, ..., N\} $ by certain quadrature rules, such as the half range Gauss-Hermite quadrature~\cite{shizgal1981gaussian} or Newton-Cotes formula. 
In the DUGKS, $\bar f$ is approximated at these discrete velocity points as $\bar f_i$, and then is used to approximate the moments, 
\begin{equation}
    \rho = \sum_{i}^{N_v}  \omega_i  \bar f_i, \quad
    \rho \boldsymbol u = \sum_{i}^{N_v} \omega_i  \boldsymbol \xi_i \bar f_i, \quad
    \rho E = \frac{1}{2} \sum _{i}^{N_v} \omega_i  \xi_i^2 \bar f_i, \quad
\end{equation}
where $N_v$ is the number of discrete velocities, $\omega_i$ is the quadrature weights. 

The time step $\Delta t$ in DUGKS is determined by the Courant-Friedrichs-Lewy (CFL) condition, 
\begin{equation}
    \Delta t = \eta \frac{\Delta x}{|\boldsymbol u|_{max} + |\boldsymbol \xi_i|_{max}},
\end{equation}
where $\eta$ is the CFL number, and $\Delta x$ is the minimum mesh size, $|\boldsymbol u|_{max}$ is the maximum flow velocity, and $|\boldsymbol \xi_i|_{max}$ is the maximum discrete velocity. 

%Additionally, for the simulation of steady flow in this paper, the convergence criterion is defined by
%\begin{equation}
 %   E(t) = \frac{\sum |\boldsymbol W(t) - \boldsymbol W(t-1000\Delta t)|}{\sum |\boldsymbol W(t)|}<10^{-6}, \quad   \text{for} \quad \boldsymbol W \in \{\rho, \boldsymbol u, T\},
%\end{equation}
%where the summations are taken over all cells.

\subsection{Isothermal Couette flow} \label{Couette flow}
\begin{figure}[htbp]
\centering
\includegraphics[width=0.6\textwidth]{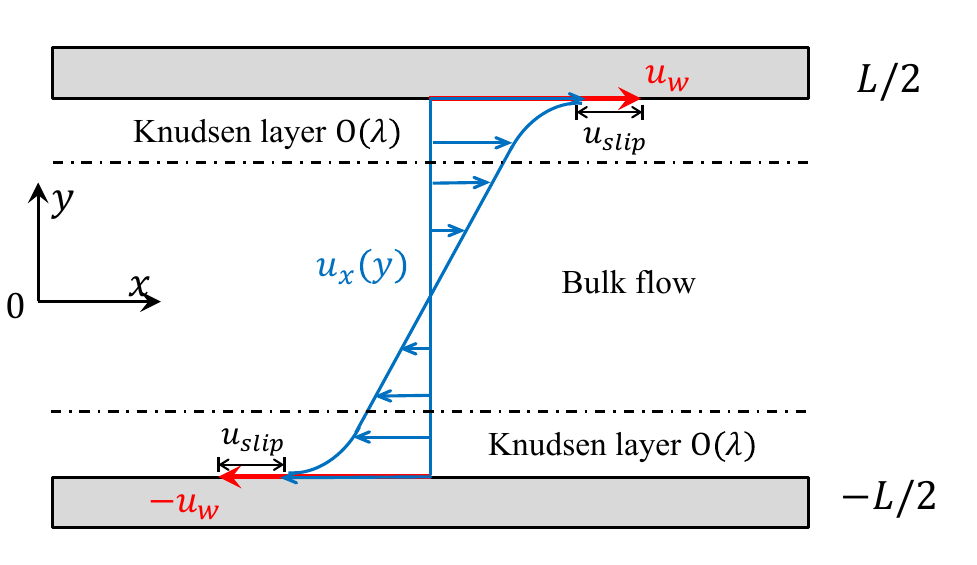}
\caption{
 Schematic of typical velocity profile for the isothermal Couette flow with slip effect. The velocity exhibits a linear spatial profile in the bulk flow and a nonlinear profile near the wall in the Knudsen layer.
}\label{FIG_Couette}
\end{figure}

The isothermal Couette flow is performed to investigate velocity slip phenomena in different flow regimes. As shown in Fig.~\ref{FIG_Couette}, the flow is confined between two infinite parallel plates located at $y=\pm L/2$ and moving with velocities $\pm u_w$ in the $x$ direction, respectively. The velocity profile is linear in the bulk flow, while the profile in the Knudsen layer near the wall is nonlinear due to non-equilibrium effects~\cite{zhang2012review}. The velocity slip $ u_{slip} = u(\boldsymbol x_w) - u_w$ is the relative velocity between the wall velocity and gas velocity at the wall. Both plates are maintained at a fixed temperature $T_0$. In this problem, the plate velocities are assumed to be much smaller than the characteristic molecular velocity $v_0=\sqrt{2RT_0}$, i.e., $u_w = 0.1 v_0$. Periodic boundary conditions are applied in the flow direction, and both the original and multiscale DMBCs are imposed on the plates. 

In our simulation, the physical space is divided into $N_x = 2 $ cells in the $x$ direction and $N_y = 10$ or $ 100$ uniform cells in the $y$ direction. The velocity space is discretized through half-range Gauss-Hermite quadrature~\cite{shizgal1981gaussian},  adopting $8 \times 8$ velocity points for $\text{Kn} \leq 0.1$ and  $28 \times 28$ velocity points for $\text{Kn} > 0.1$. The CFL number is set to be 0.5 in all cases. The flow field is assumed to be steady when the maximum relative change of the velocity field in two successive steps is less than $10^{-10}$. 

\begin{figure}[htbp]
\centering
\includegraphics[width=0.5\textwidth]{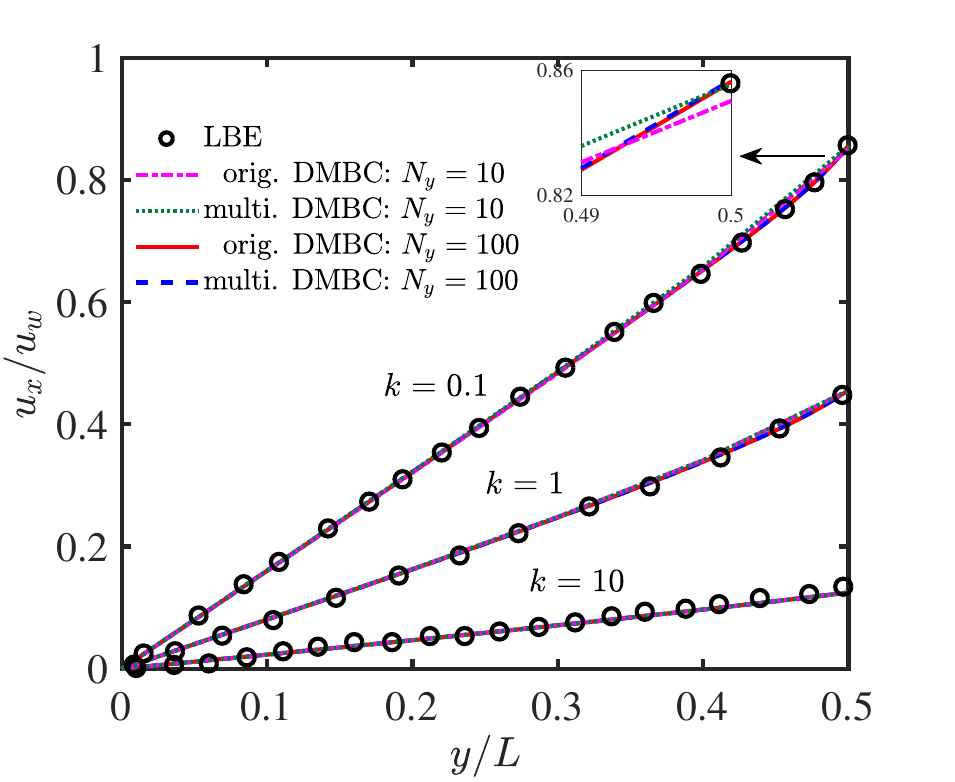}
\caption{
Velocity profiles of the Couette flow at different Knudsen numbers $[k=(\sqrt{\pi}/2)\text{Kn}]$ obtained from different DMBCs (the fully diffuse-reflection condition, i.e., $\alpha = 1$) with $N_y = 10$ and $N_y = 100$ uniform cells. In this and the subsequent plots, "orig. DMBC" denotes results obtained using the DUGKS with the original DMBC, while "multi. DMBC" denotes results from the DUGKS employing the multiscale DMBC.
The LBE data are from Ref.~\cite{sone1990numerical}. 
}\label{FIG_ComparaRarefied_Couetteflow}
\end{figure}

\begin{figure}[htbp]
\centering
\subfigure[]{\includegraphics[width=0.32\textwidth]{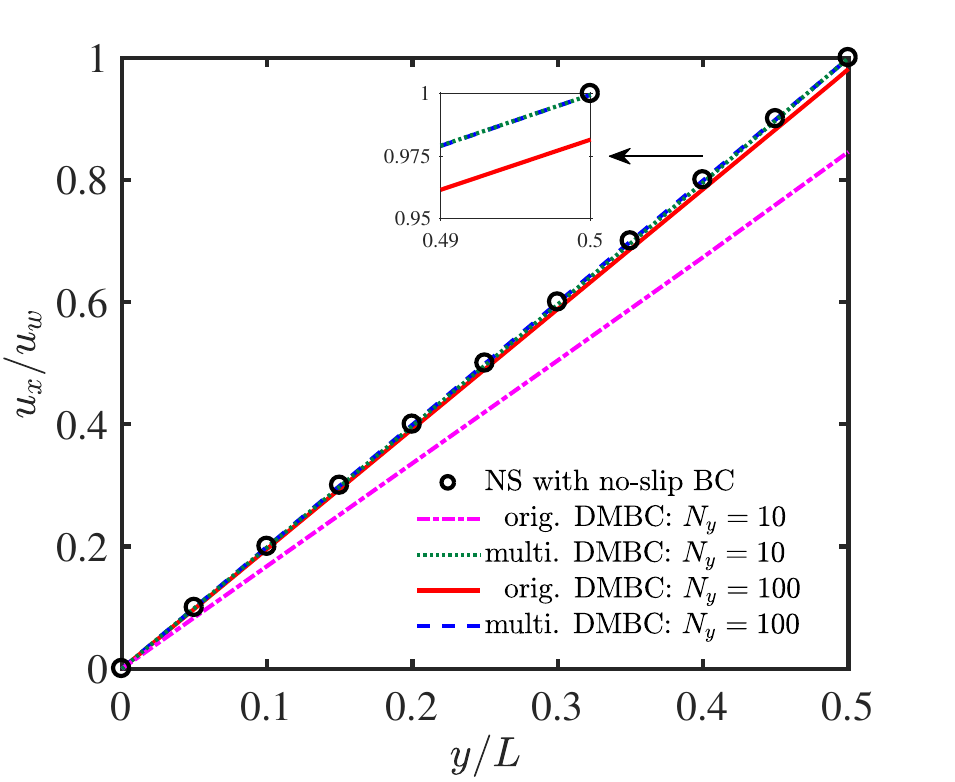}\label{sub_u_TMAC0.1}}
\subfigure[]{\includegraphics[width=0.32\textwidth]{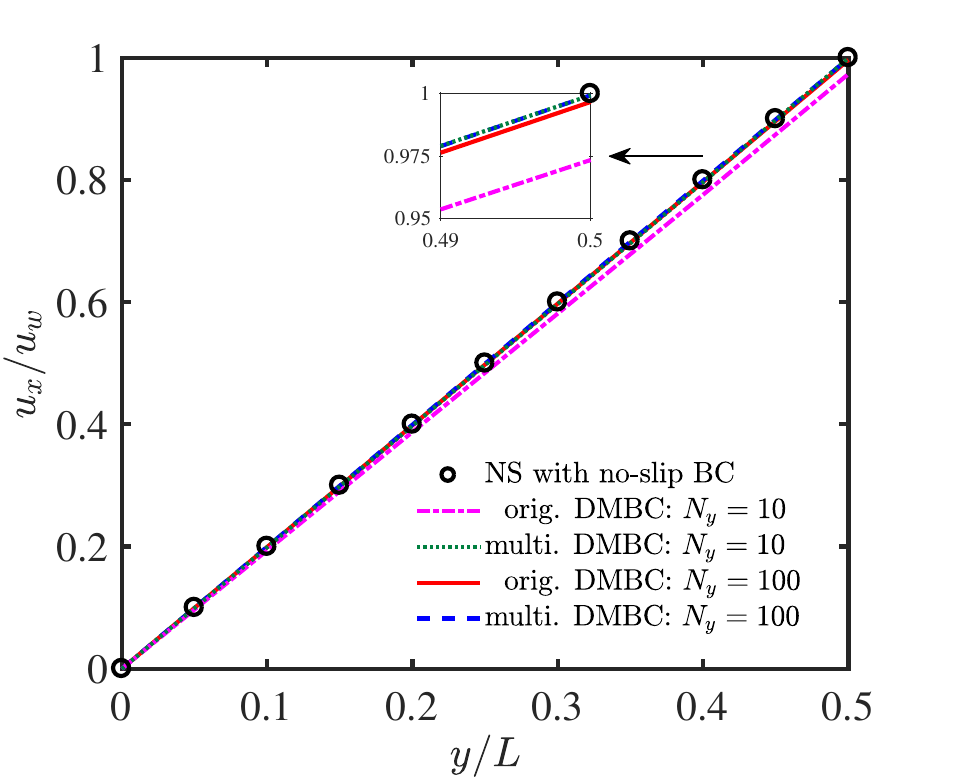}\label{sub_u_TMAC0.5}}
\subfigure[]{\includegraphics[width=0.32\textwidth]
{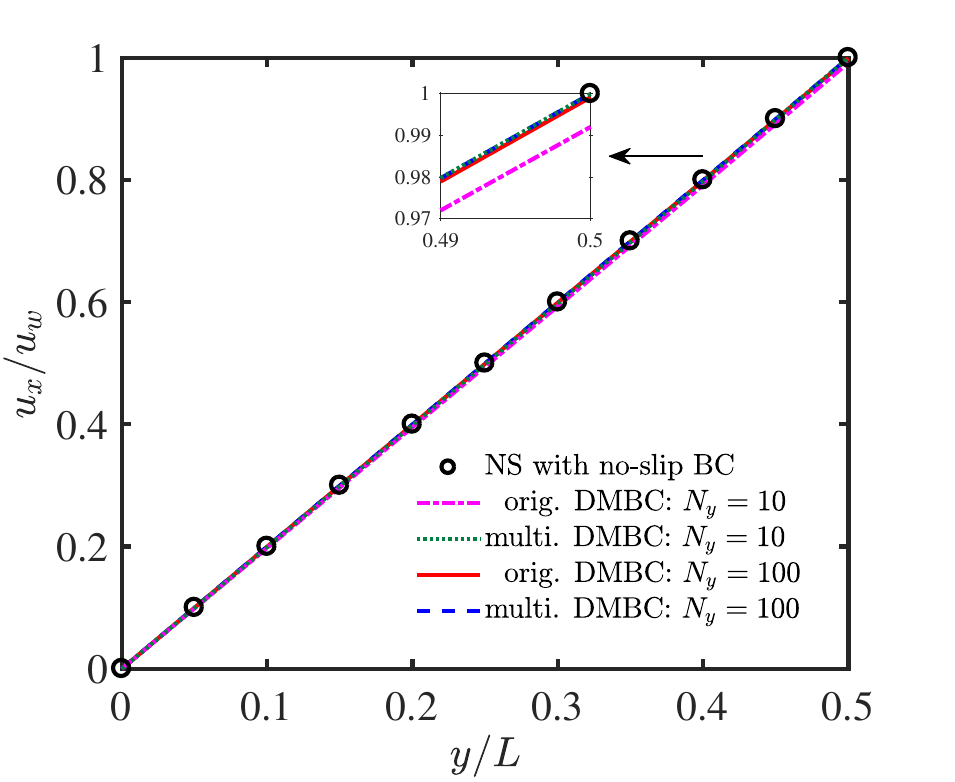}\label{sub_u_TMAC1}}
\caption{Velocity profiles of the Couette flow in the continuum flow regime ($k = 1.0\times 10^{-5}$) obtained from different DMBCs with $N_y = 10$ and $N_y = 100$ uniform cells for different accommodation coefficient: (a) $\alpha = 0.1$, (b) $\alpha = 0.5$, and (c) $\alpha = 1.0$.}
\label{FIG:Couetteflow_u_differentTMAC}
\end{figure}

Initially, we consider the case of fully diffuse-reflection conditions ($\alpha = 1$) for validation.
The normalized velocity profiles with $k=(\sqrt{\pi}/2)\text{Kn} =0.1$, $1$, and $10$, along with the benchmark data obtained by the linearized Boltzmann equation (LBE)~\cite{sone1990numerical}, are shown in Fig.~\ref{FIG_ComparaRarefied_Couetteflow} . 
As we can see, the results of the original and multiscale DMBCs show good agreement with the benchmark data in all cases. In particular, both DMBCs can give adequately accurate results with just $N_y=10$ cells. However, for the results of the original DMBC with $N_y=10$ cells at $k=0.1$, there is a minor error in the Knudsen layer. The slip velocity at the wall predicted by the multiscale DMBC is more accurate than the original one.  Furthermore, the normalized velocity profiles for different accommodation coefficients $\alpha = 0.1, 0.5$, and $1.0$ in the continuum flow regime ($k=1.0 \times 10^{-5}$) are presented in Fig.~\ref{FIG:Couetteflow_u_differentTMAC}, accompanied the NS solution with no-slip BCs. It is noteworthy that the slip velocity, proportional to $k$ (or $\text{Kn}$) and approaching zero as decreases, justifies the use of the NS equations with no-slip conditions for this comparison. It can be observed that the multiscale DMBC provides sufficiently accurate results with only 10 cells, showing good agreement with both the results using on 100 cells and the reference solution. However, the results with the original DMBC exhibit significant errors, which become increasingly pronounced as the value of $\alpha$ decreases. This is attributed to the fact that the original DMBC produces a spurious velocity slip in continuum flow, which is directly proportional to both the coefficient $(2-\alpha)/\alpha$ and time step (or cell size), as deduced in Sec.~\ref{Continuum limit}.

\begin{figure}[htbp]
\centering
\subfigure[]{\includegraphics[width=0.32\textwidth]{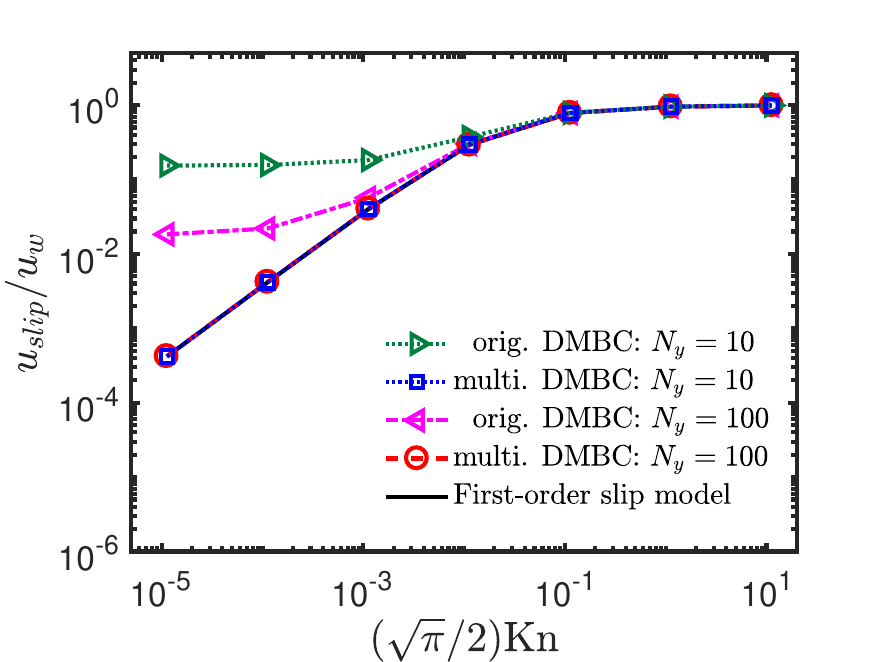}\label{sub_slip_TMAC0.1}}
\subfigure[]{\includegraphics[width=0.32\textwidth]{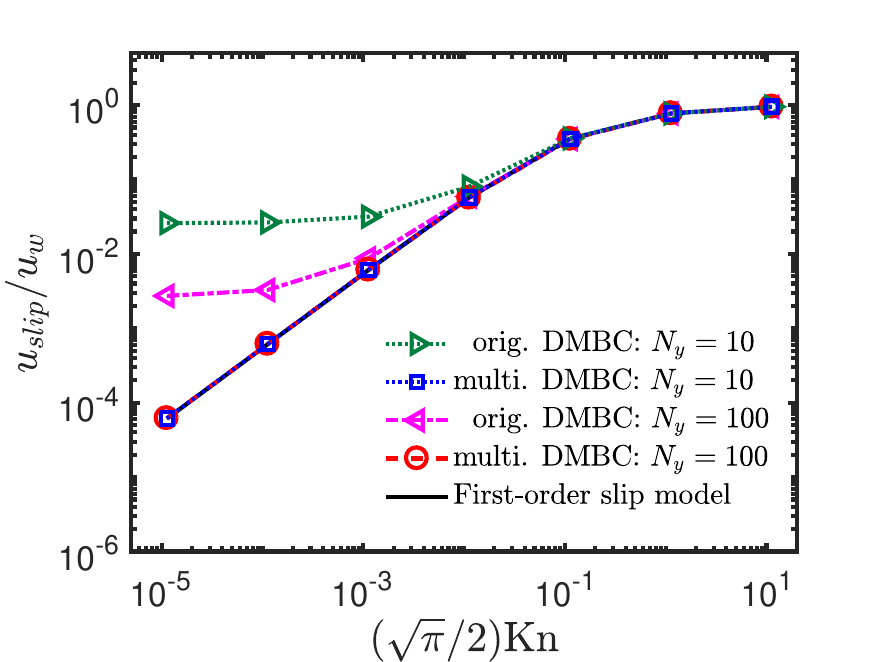}\label{sub_slip_TMAC0.5}}
\subfigure[]{\includegraphics[width=0.32\textwidth]{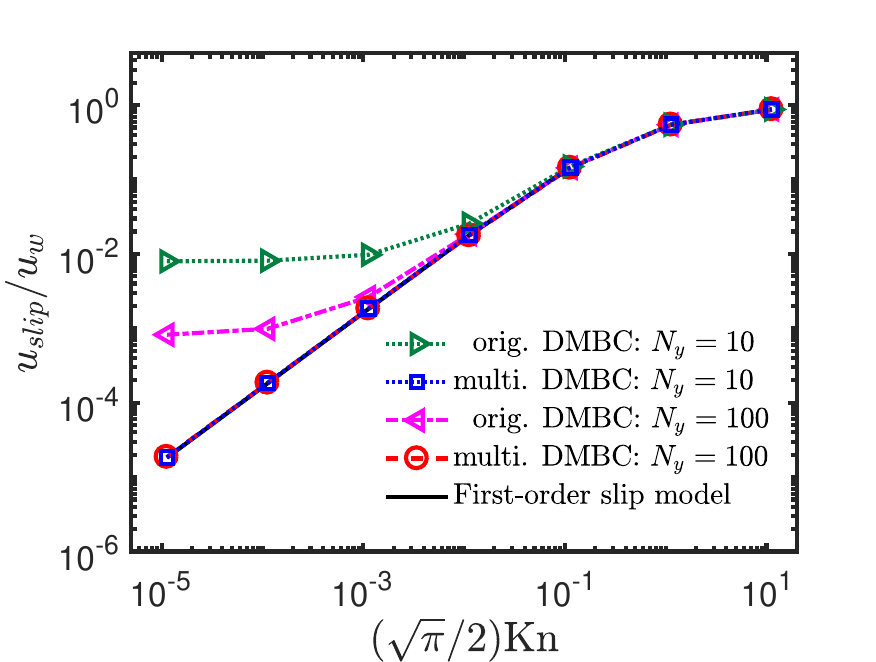}\label{sub_slip_TMAC1.0}}
\caption{Normalized velocity slip profiles of Couette flow versus $k$ with $N_y = 10$ and $N_y = 100$ uniform cells for different accommodation coefficient: (a) $\alpha = 0.1$, (b) $\alpha = 0.5$, and (c) $\alpha = 1.0$.}
\label{FIG:Couetteflow_slip_differentTMAC}
\end{figure}

Moreover, the difference in velocity slip between the original and multiscale DMBCs will be compared quantitatively. 
It is found that the slip velocities of the upper and lower plates are almost the identical, and the normalized slip velocity profiles of the upper plate versus $k$ for different accommodation coefficient $\alpha$ are shown in Fig.~\ref{FIG:Couetteflow_slip_differentTMAC}. Here, the ratio of $\tau$ to $h$ is $\tau/h = 41 \times N_y \times \text{Kn} $ for different Kn and cell numbers. In addition, the first-order slip model is included for comparison, which is valid for both continuum and slip regimes (i.e. $\text{Kn}< 0.1$)~\cite{loyalka1975some},
\begin{equation}
    u_{slip} = \frac{2-\alpha}{\alpha}\left( 1- 0.1817\alpha \right) \text{Kn} \frac{\partial u_x (\boldsymbol x_w)}{\partial (y/L)}, 
\end{equation}
where the local velocity gradient can be computed using the relation ${\partial u_x (\boldsymbol x_w)}/{\partial (y/L)} = {\sigma_{xy}L}/{\mu}$ and the shear stress obtained by Eq.~\eqref{eq:high order macro}. 
%The values of shear stress $\sigma_{xy}$ corresponding to different Kn and accommodation coefficient $\alpha$ are listed in Table 
As we seen in Fig.~\ref{FIG:Couetteflow_slip_differentTMAC}, the results of the original and multiscale DMBCs with $N_y = 10$ and $N_y = 100$ cells are in excellent agreement in rarefied regimes ($k>0.1$) for all $\alpha$. However, a clear difference between the two DMBCs can be observed in continuum and slip regimes (i.e., $k< 0.1$). The multiscale DMBC with 10 cells can accurately capture the slip phenomenon compared to the first-order slip model, i.e., the slip velocity decreases linearly as $k$ decreases. However, it can be observed that $u_{slip}$ with the original DMBC first decreases and then remains constant with decreasing $k$ for all $\alpha$.  Particularly, the spurious slip magnitude at $\alpha = 0.1$ is significantly greater than the results for $\alpha = 0.5$ and $\alpha = 1.0$.
These numerical results also demonstrate the arguments for continuum limit in Sec.~\ref{Continuum limit}. 

\begin{figure}[htbp]
\centering
{\includegraphics[width=0.5\textwidth]{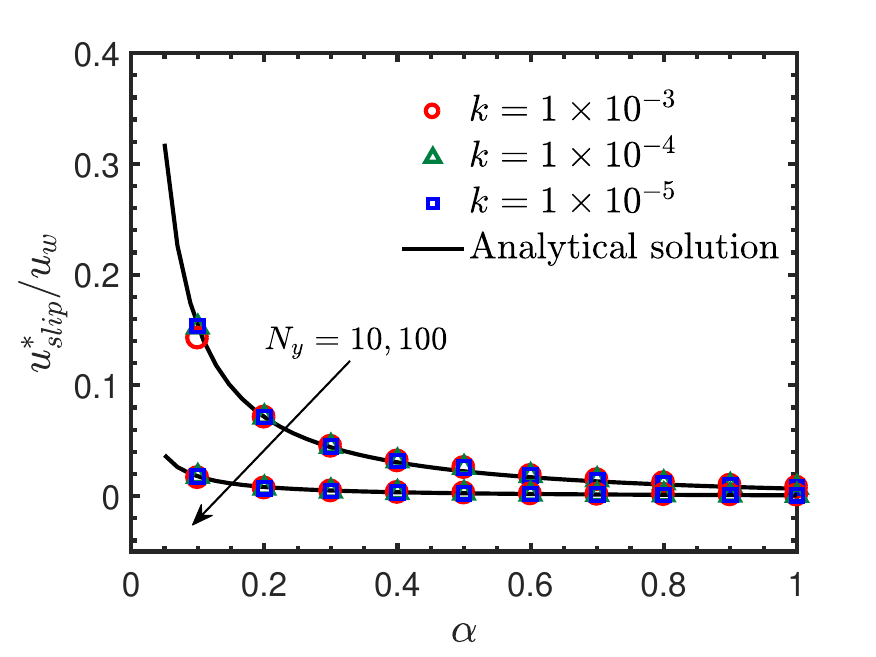}\label{FIG_delta}}
\caption{Normalized spurious velocity slip profiles of Couette flow versus accommodation coefficient $\alpha$ obtained from the original DMBC with $N_y = 10$ and $N_y = 100$ uniform cells. The analytical solution is given by Eq.~\eqref{spurious velocity slip}.}
\label{velocity slip profiles versus alpha}
\end{figure}

Finally, we can derive the following expression for the spurious velocity slip following the method in Ref.~\cite{loyalka1975some},
\begin{equation} \label{spurious velocity slip}
    u^*_{slip}= \frac{2-\alpha}{\alpha}(1-0.1817\alpha)\frac{h}{L}\sqrt{\frac{\pi R T_0}{2}}\frac{\partial u_x(\boldsymbol x_w)}{\partial (y/L)},
\end{equation}
which suggests that the spurious slip depends on the accommodation coefficient $\alpha$ and time step (or cell size).
Figure~\ref{velocity slip profiles versus alpha} shows the normalized spurious slip velocity obtained by the original DMBC versus $\alpha$ for $N_y = 10$ and $N_y = 100$ uniform cells, along with the analytical solution~\eqref{spurious velocity slip} in continuum regimes (i.e., $k<0.001$). Clearly, the analytical solution~\eqref{spurious velocity slip} can accurately predict the spurious slip velocity produced by the original DMBC. The magnitude of the non-physical spurious slip decreases with increasing $\alpha$ and cell number.

\subsection{Fourier flow}
\begin{figure}[htbp]
\centering
\includegraphics[width=0.6\textwidth]{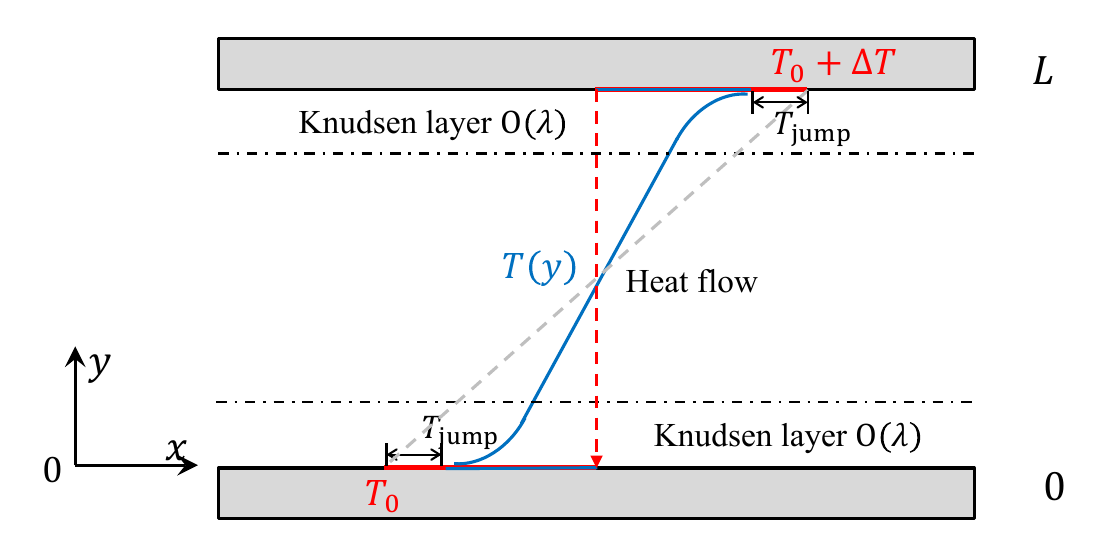}
\caption{
 Schematic of typical temperature profile for the Fourier flow with temperature jump. The temperature exhibits a linear spatial profile in the bulk flow and a nonlinear profile near the wall in the Knudsen layer.
}\label{FIG_Fourier}
\end{figure}

We now investigate the temperature jump of the Fourier flow. The geometric of this problem is similar to the Couette flow but with stationary plates. The temperature jump $T_{jump}$ is the difference between wall temperature and gas temperature at the wall. As shown in Fig.~\ref{FIG_Fourier}, the temperatures
at the upper and lower plates are set to be $T_0 +\Delta T$  and $T_0 $ with $\Delta T = 0.1 T_0$, respectively. The rest of the simulation conditions are the same as the Couette flow. The steady is reached as the maximum relative change of the heat flux field in two successive steps is less than $10^{-10}$. 

\begin{figure}[htbp]
\centering
\subfigure[]{\includegraphics[width=0.32\textwidth]{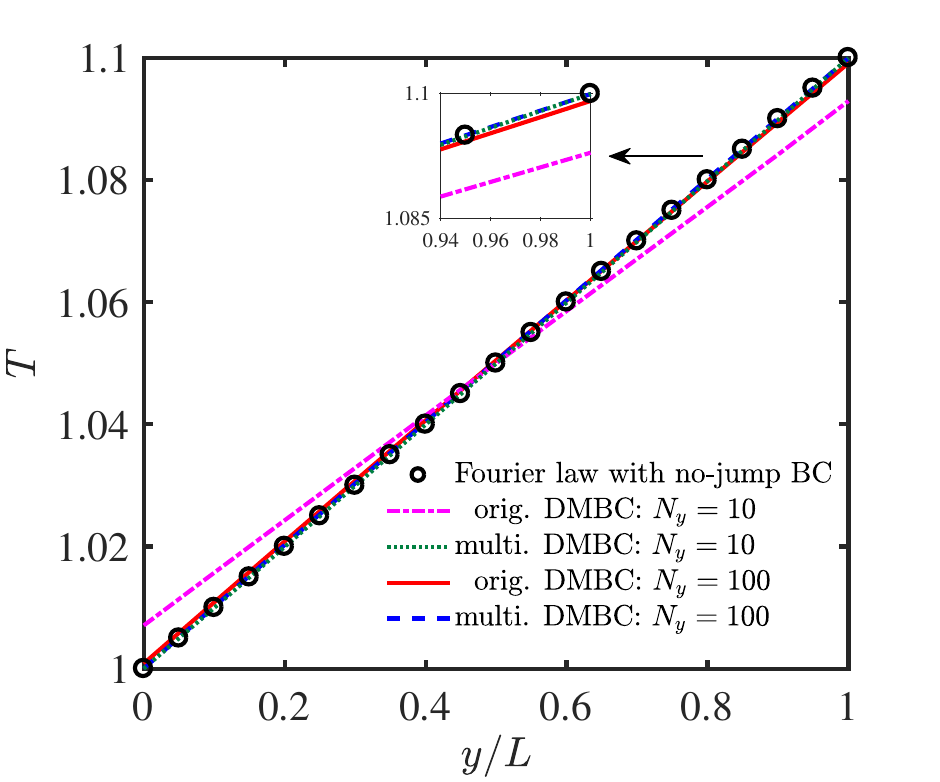}\label{sub_T_TMAC0.1}}
\subfigure[]{\includegraphics[width=0.32\textwidth]{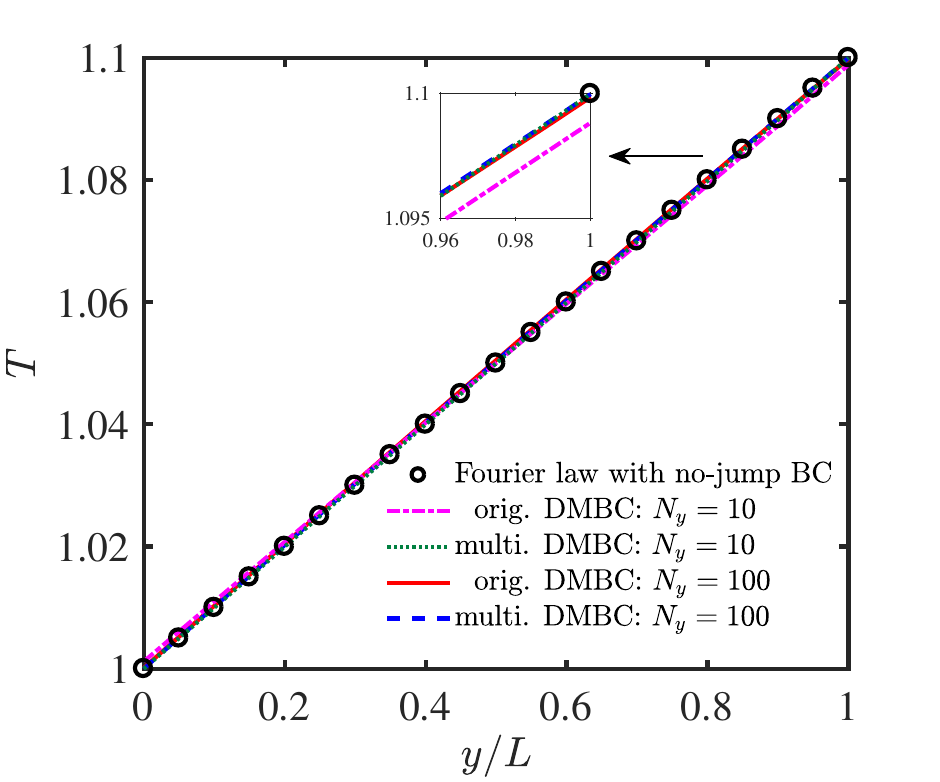}\label{sub_T_TMAC0.5}}
\subfigure[]{\includegraphics[width=0.32\textwidth]
{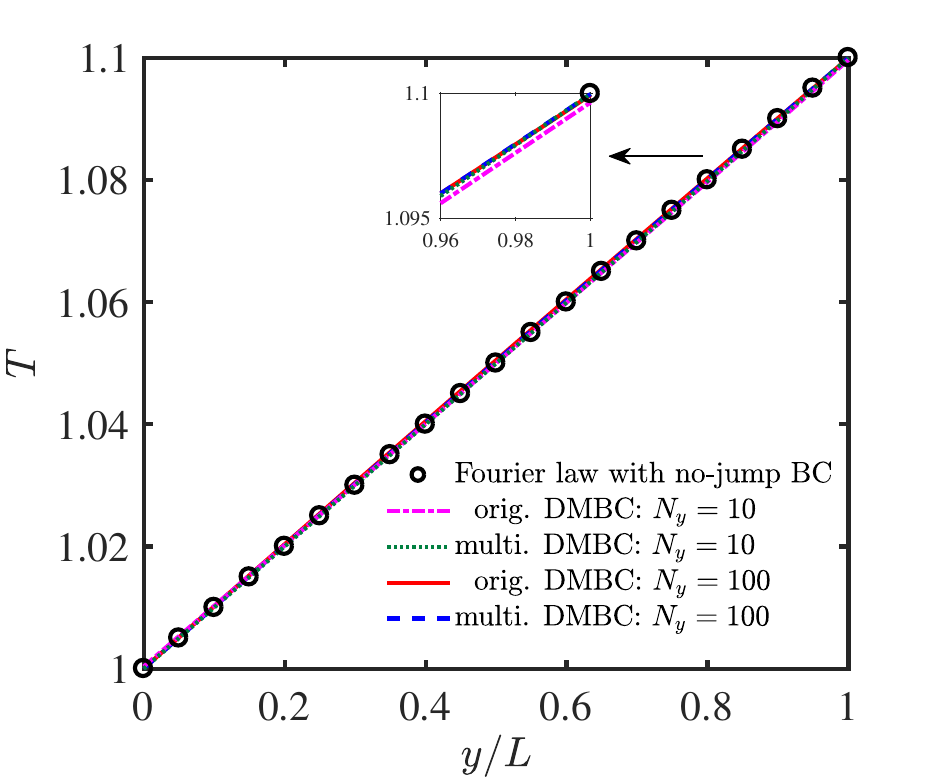}\label{sub_T_TMAC1}}
\caption{Normalized temperature profiles of the Fourier flow in the continuum flow regime ($k = 1.0\times 10^{-5}$) obtained from different DMBCs with $N_y = 10$ and $N_y = 100$ uniform cells for different accommodation coefficient: (a) $\alpha = 0.1$, (b) $\alpha = 0.5$, and (c) $\alpha = 1.0$.}
\label{FIG:Fourierflow_T_differentTMAC}
\end{figure}

The normalized temperature and density profiles obtained from both DMBCs in the rarefied flow regime ($ \text{Kn} = 0.01$, $ 0.1$, $1$, and $10$) demonstrate good agreement with the reference solution of the Boltzmann equation~\cite{wu2013deterministic}, and are not presented here. In the continuum flow regime ($\text{Kn}=1.0\times 10^{-5}$), the normalized temperature profiles for different accommodation coefficients $\alpha=0.1$, $0.5$, and $1.0$ are shown in Fig.~\ref{FIG:Fourierflow_T_differentTMAC}, in which the Fourier law with no-jump BCs serves as the reference solution. Note that the temperature jump justifies the use of the Fourier law with no-jump conditions. Similar to slip velocity, the temperature jump at the wall predicted by the multiscale DMBC is more accurate than the original one. Particularly, the multiscale DMBC provides sufficiently accurate results with coarser mesh, while the results predicted using the original DMBC exhibit significant errors, which become increasingly pronounced for small $\alpha$. This can be attributed to the fact that the original DMBC produces a spurious temperature jump in continuum flow, which is directly proportional to both the coefficient $(2-\alpha)/\alpha$ and time step (or cell size), as deduced in Sec.~\ref{Continuum limit}.

\begin{figure}[htbp]
\centering
\subfigure[]{\includegraphics[width=0.32\textwidth]{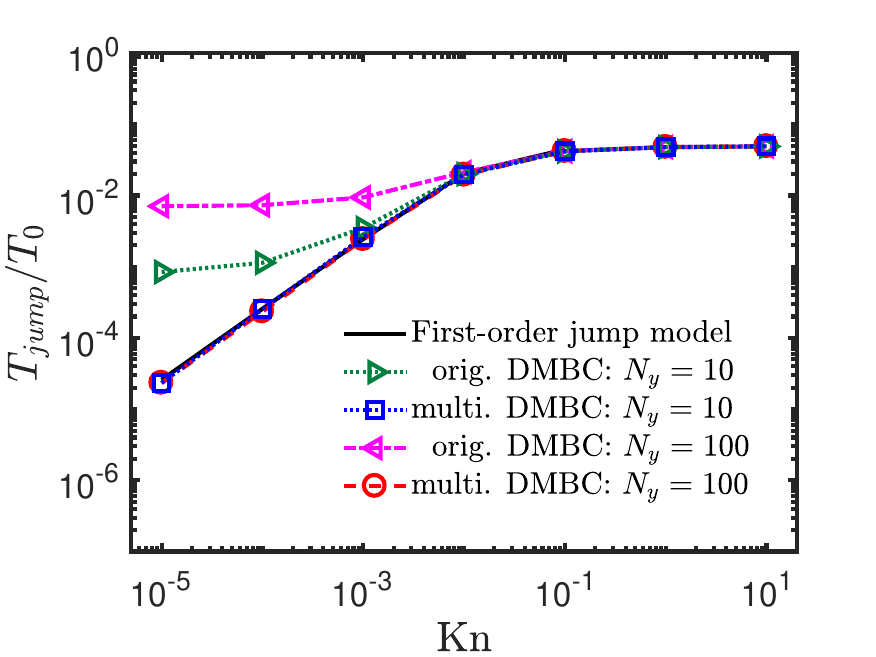}\label{sub_jump_TMAC0.1}}
\subfigure[]{\includegraphics[width=0.32\textwidth]{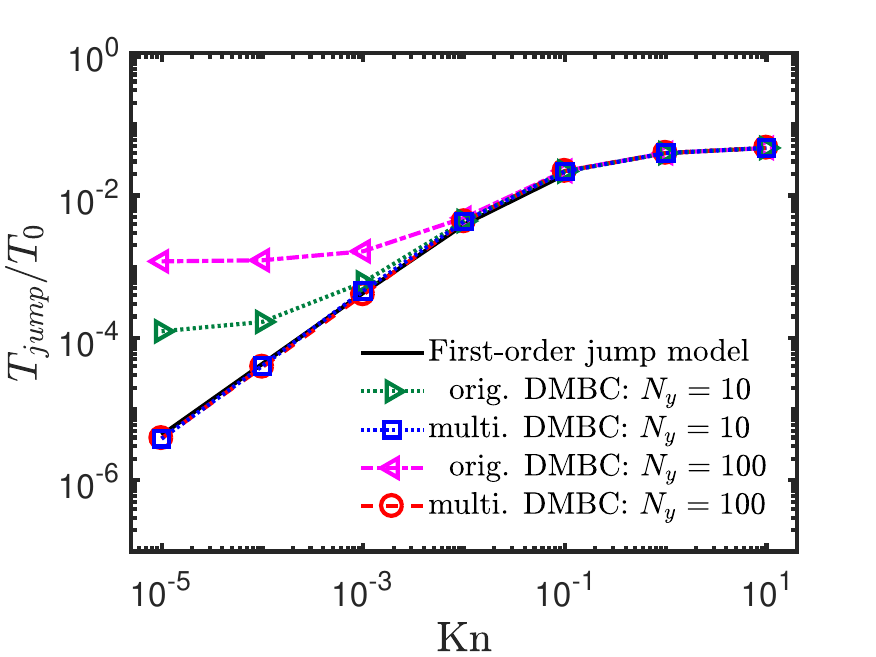}\label{sub_jump_TMAC0.5}}
\subfigure[]{\includegraphics[width=0.32\textwidth]{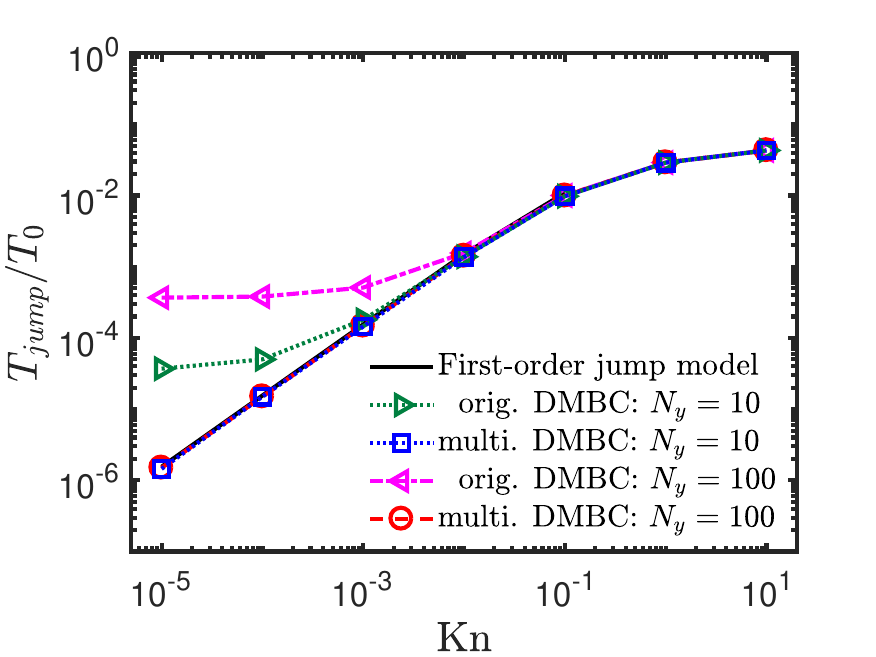}\label{sub_jump_TMAC1.0}}
\caption{Normalized temperature jump profiles of Fourier flow versus Kn with $N_y = 10$ and $N_y = 100$ uniform cells for different accommodation coefficient: (a) $\alpha = 0.1$, (b) $\alpha = 0.5$, and (c) $\alpha = 1.0$.}
\label{FIG:Fourier_jump_differentTMAC}
\end{figure}

Further quantitative comparisons of the temperature jump predicted by both DMBCs are investigated. It is found that the qualitative results of the temperature jump are the same for the upper and lower plates, and the normalized temperature jump profiles of the lower plate versus Kn for different accommodation coefficients $\alpha$ are shown in Fig.\ref{FIG:Fourier_jump_differentTMAC}. Additionally, the first-order temperature jump model is included for comparison, which is valid for both continuum and slip regimes (i.e., $\text{Kn}<0.1$)~\cite{1968Momentum},
\begin{equation}
    T_{jump} = \frac{15}{8}\frac{2-\alpha}{\alpha}\left(1+0.1621 \alpha \right) \text{Kn}\frac{\partial T(\boldsymbol x_w)}{\partial (y/L)},
\end{equation}
for the lower plates. The local temperature gradient can be computed using the relation ${\partial T (\boldsymbol x_w)}/{\partial (y/L)} = {q_y L}/{\kappa}$, where the heat flux is obtained from Eq.~\eqref{eq:high order macro} and the thermal conductivity is given by $\kappa=5/2R\mu$. Similar to the Couette flow, the results of both DMBCs with $N_y = 10$ and $N_y = 100$ cells are in good agreement in rarefied regimes ($\text{Kn}>0.1$) for all values of $\alpha$. However, a clear difference between the two DMBCs can be observed in continuum and slip regimes (i.e., $\text{Kn}< 0.1$). The multiscale DMBC with 10 cells can capture the temperature jump phenomenon compared to the first-order jump model. However, the temperature jump of the original DMBC first decreases and then remains constant with decreasing Kn for all $\alpha$. 

\begin{figure}[htbp]
\centering
{\includegraphics[width=0.5\textwidth]{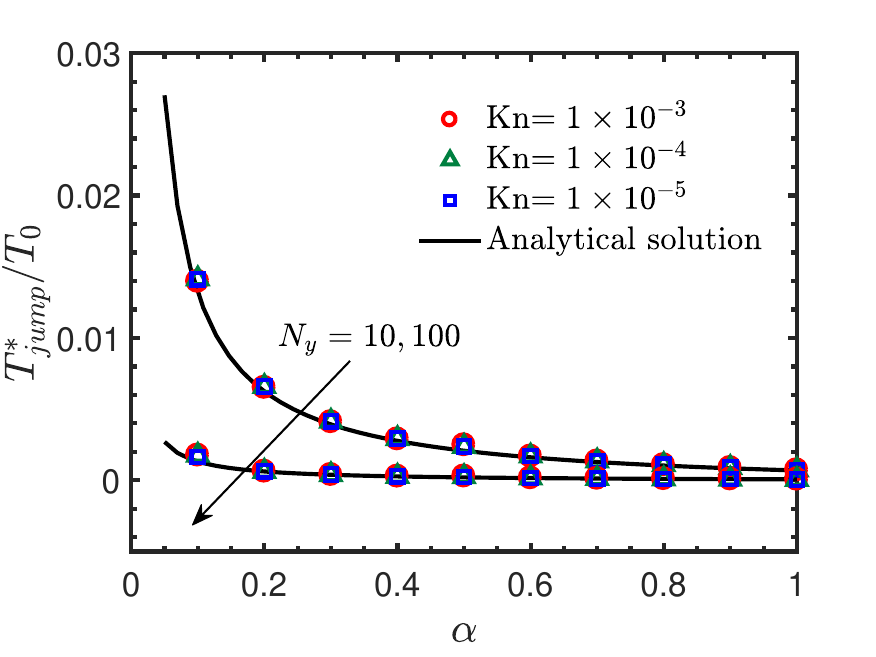}}
\caption{Normalized spurious temperature jump profiles of Fourier flow versus accommodation coefficient $\alpha$ obtained from the original DMBC with $N_y = 10$ and $N_y = 100$ uniform cells. The analytical solution is given by Eq.~\eqref{spurious velocity slip}.}
\label{temperature jump profiles versus alpha}
\end{figure}

These numerical results also demonstrate the arguments in Sec.~\ref{Continuum limit}. Specifically, we can derive the following expression for the spurious temperature jump produced by the original DMBC following the approach in Ref.~\cite{1968Momentum},
\begin{equation}\label{spurious T jump}
    T^*_{jump} = \frac{15}{8}\frac{2-\alpha}{\alpha}\left(1+0.1621 \alpha \right) \frac{h}{L}\sqrt{\frac{\pi RT}{2}}\frac{\partial T(\boldsymbol x_w)}{\partial (y/L)}.
\end{equation}
Figure~\ref{temperature jump profiles versus alpha} shows the normalized temperature jump obtained by the original DMBC versus $\alpha$ for $N_y = 10$ and $N_y = 100$ uniform cells, along with the analytical solution~\eqref{spurious T jump}  in continuum regimes (i.e., $\text{Kn}<0.001$). Obviously, the numerical results are in good agreement with the analytical solution. Therefore, Eq.~\eqref{spurious T jump} can be used to predict the spurious temperature jump generated by the original DMBC. The magnitude of the non-physical spurious temperature jump decreases with increasing $\alpha$ and $N_y$.

\subsection{Lid-driven cavity flow}
\begin{figure}[htbp]
\centering
\includegraphics[width=0.5\textwidth]{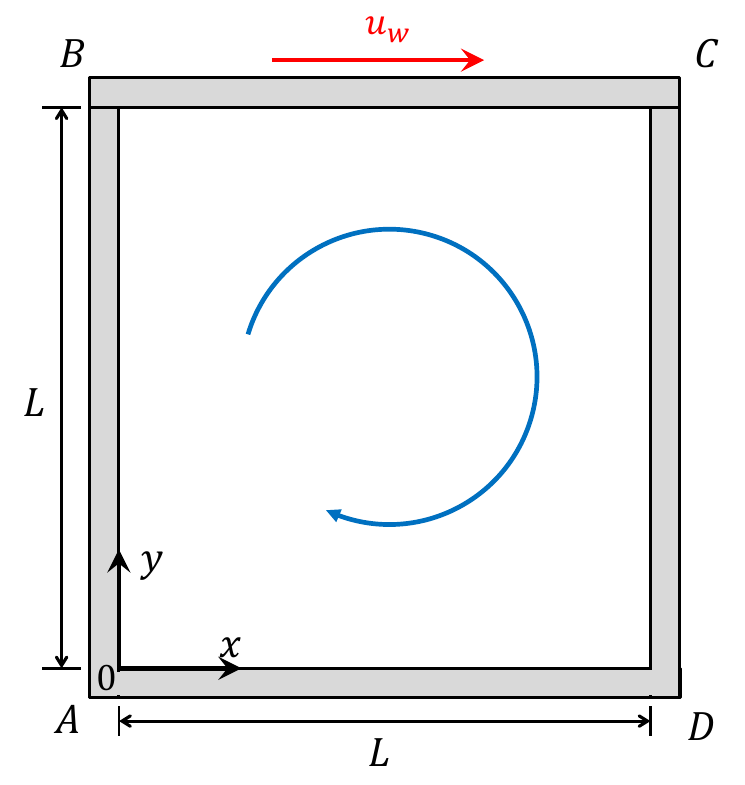}
\caption{
 Schematic of two-dimensional lid-driven cavity flow with a length of $L$, wall temperature of $T_0$ and lid velocity of $u_w$.
}\label{FIG_Cavity}
\end{figure}

In addition to the quasi-one-dimensional flow problems, a comparative study is also performed by simulating the two-dimensional lid-driven cavity flow. As illustrated in Fig.~\ref{FIG_Cavity}, the gas is contained in a square cavity of side length $L$, with the four corners denoted by A, B, C, and D. The top wall (the lid) moving along the $x$ direction with a constant velocity $u_w$, while the other walls are at rest. All the solid walls are maintained at a uniform reference temperature $T_0$, where the original and multiscale DMBCs are used. In the simulation,the driven velocity is set to be $0.16\sqrt{\gamma RT_0}$ so that the Mach number is small. Uniform meshes with cell number $N \times N$ are used in the physical space. The velocity space is discretized via half-range Gauss-Hermit quadrature for $\text{Re}=1000$ with $8 \times 8$ velocity points and for $\text{Kn}=0.075$ with $28 \times 28$ velocity points \cite{shizgal1981gaussian}. The Newton–Cotes
quadrature with  $101 \times 101$ velocity points distributed uniformly
in $[-4\sqrt{2RT}, 4\sqrt{2RT}] \times [-4\sqrt{2RT}, 4\sqrt{2RT}]$ for $\text{Kn} = 1$, and $10$.

\begin{figure}[htbp]
\centering
{\includegraphics[width=0.45\textwidth]{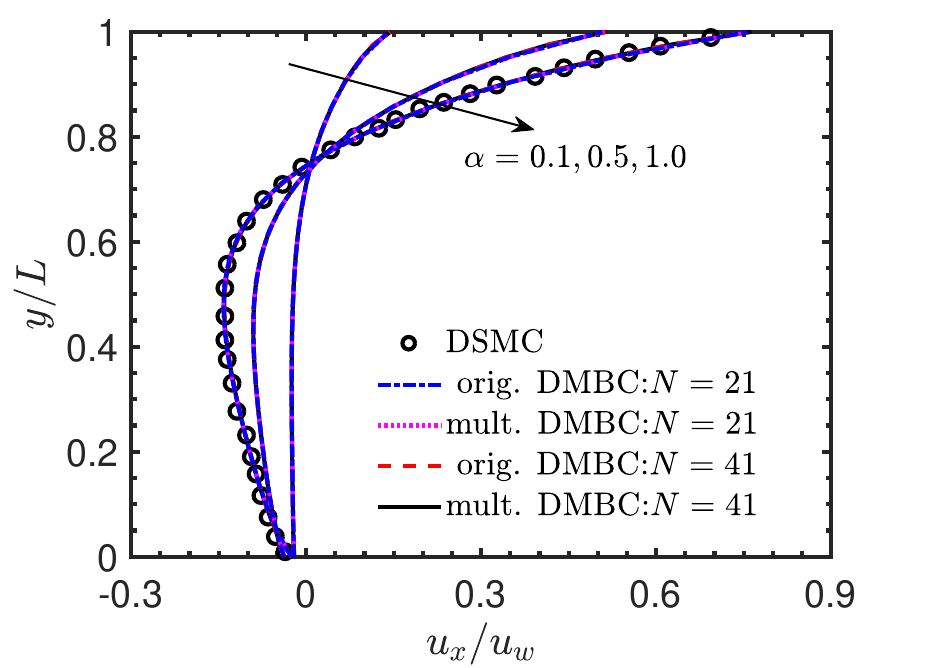}\label{FIG_Kn0075_ux}}
{\includegraphics[width=0.45\textwidth]{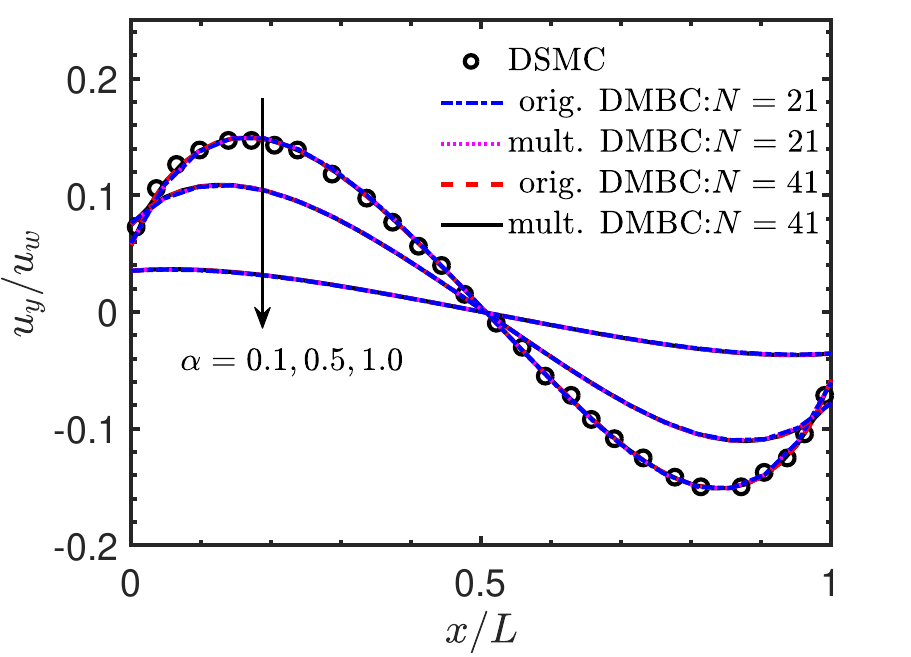}\label{FIG_Kn0075_uy}}\\
{\includegraphics[width=0.45\textwidth]{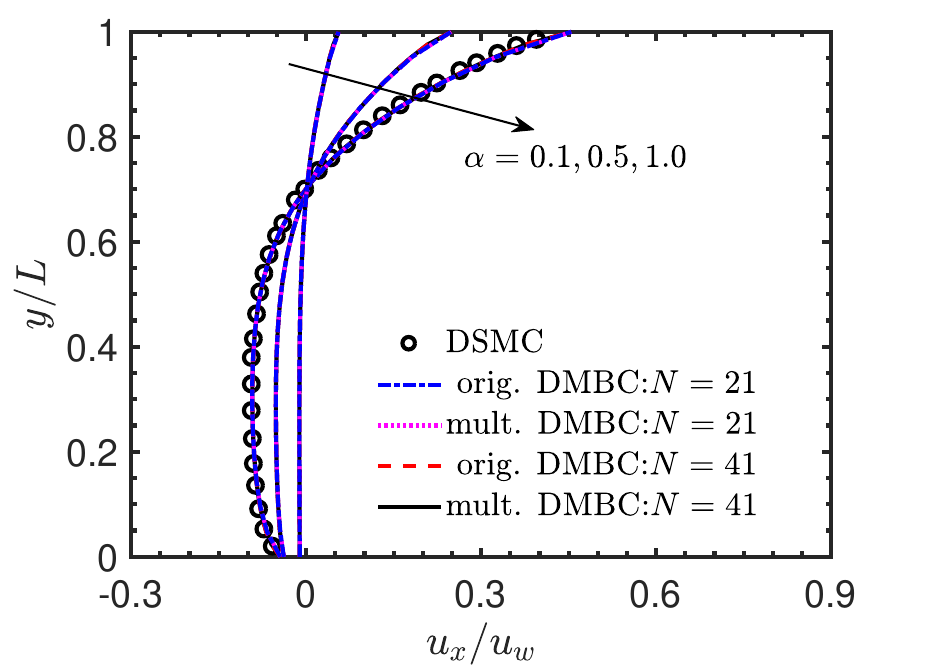}\label{FIG_Kn1_ux}}
{\includegraphics[width=0.45\textwidth]{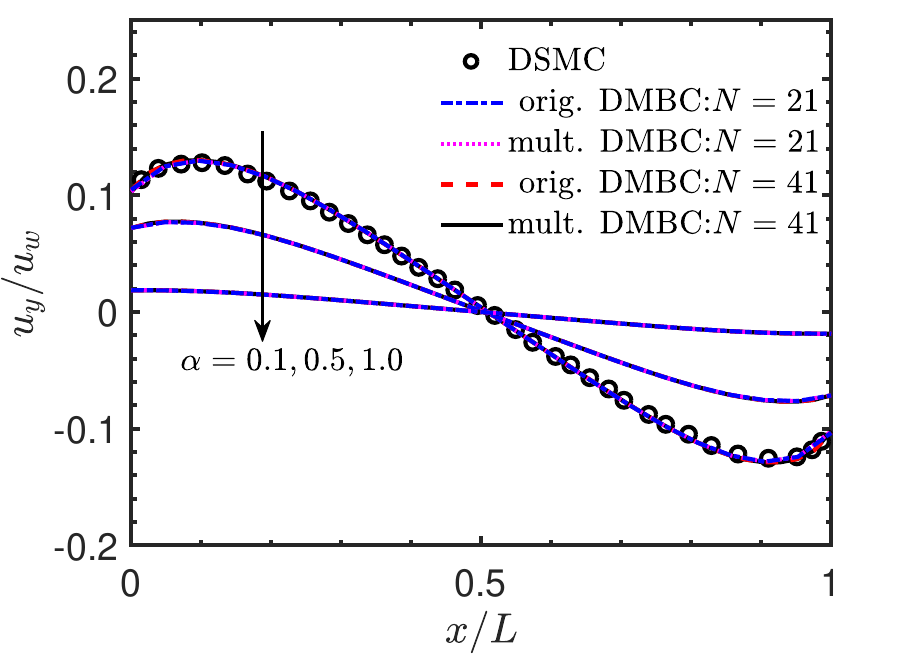}\label{FIG_Kn1_uy}}\\
{\includegraphics[width=0.45\textwidth]{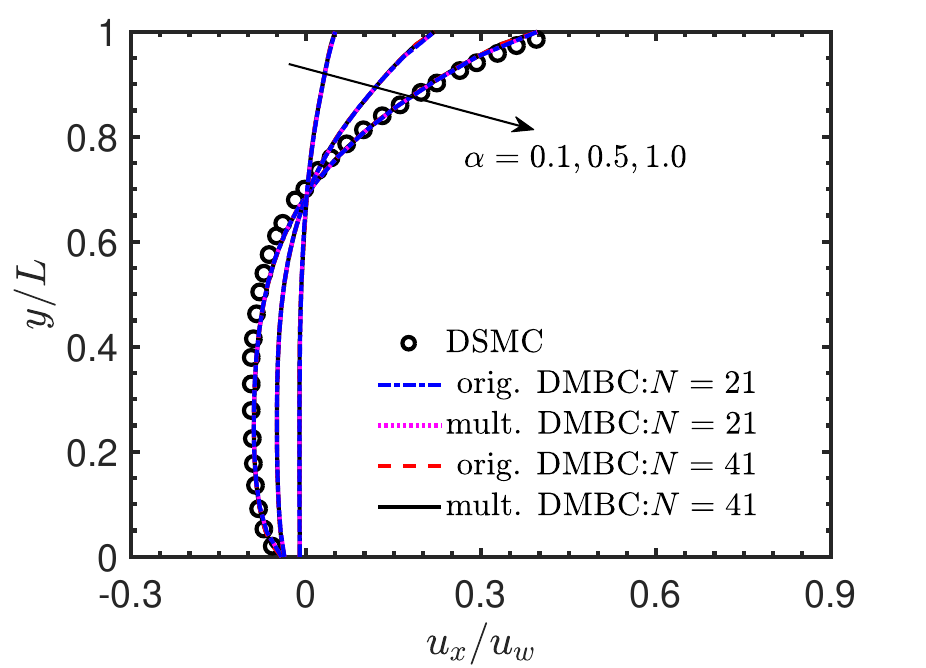}\label{FIG_Kn10_ux}}
{\includegraphics[width=0.45\textwidth]{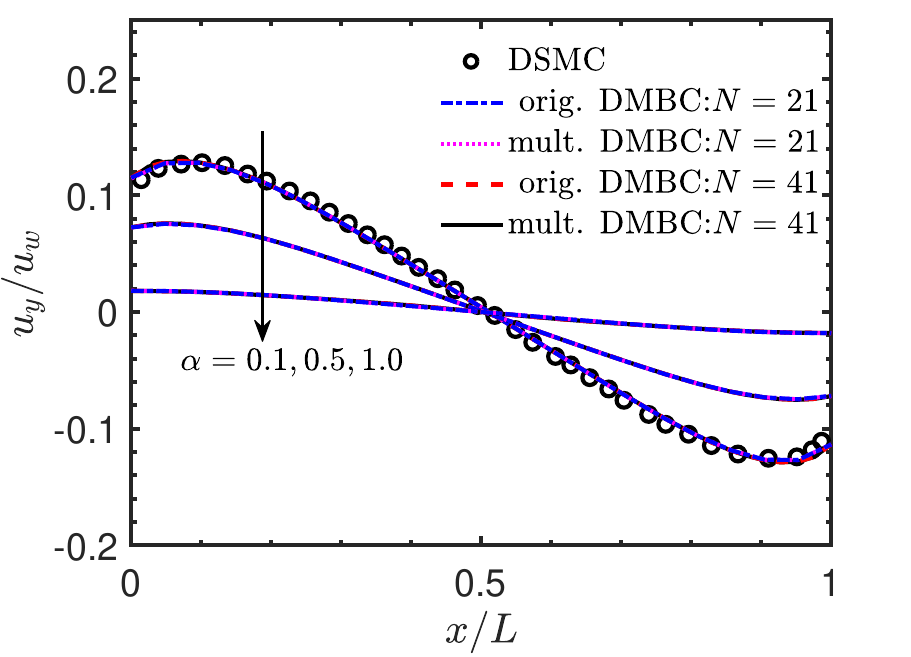}\label{FIG_Kn10_uy}}
\caption{Velocity profiles along the vertical centerline (1st column) and horizontal centerline (2nd column) of the cavity with different meshes and different accommodation coefficients for different Kn: $\text{Kn} = 0.075$ (1st row), $\text{Kn} = 0.1$ (2nd row), $\text{Kn} = 10$ (3rd row). The benchmark data are from Ref~\cite{john2011effects}.}
\label{FIG:cavity velocity Kn}
\end{figure}

\begin{figure}[htbp]
\centering
{\includegraphics[width=0.32\textwidth]{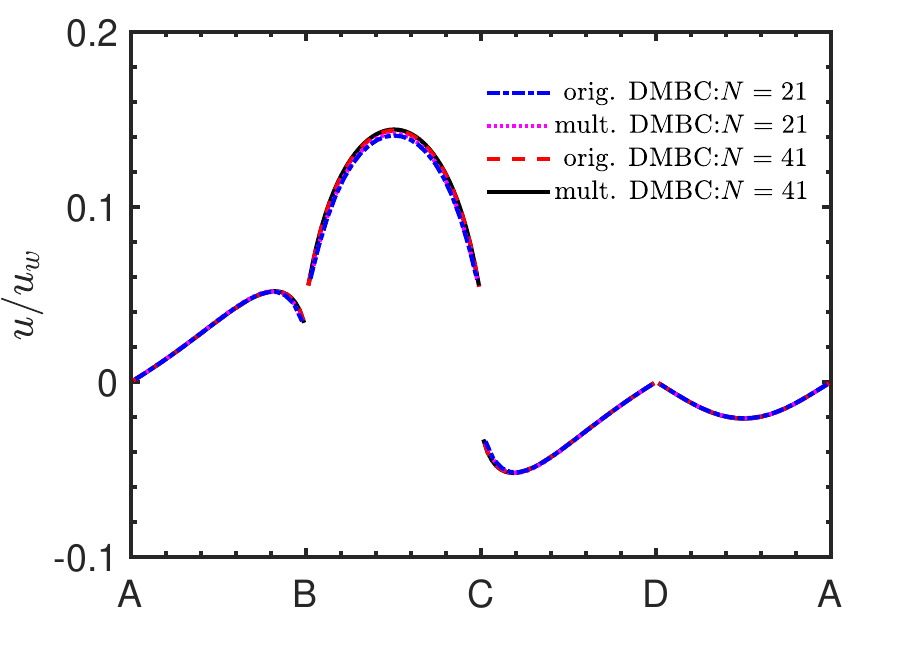}\label{FIG_Kn0075_uslip_TMAC01}}
{\includegraphics[width=0.32\textwidth]{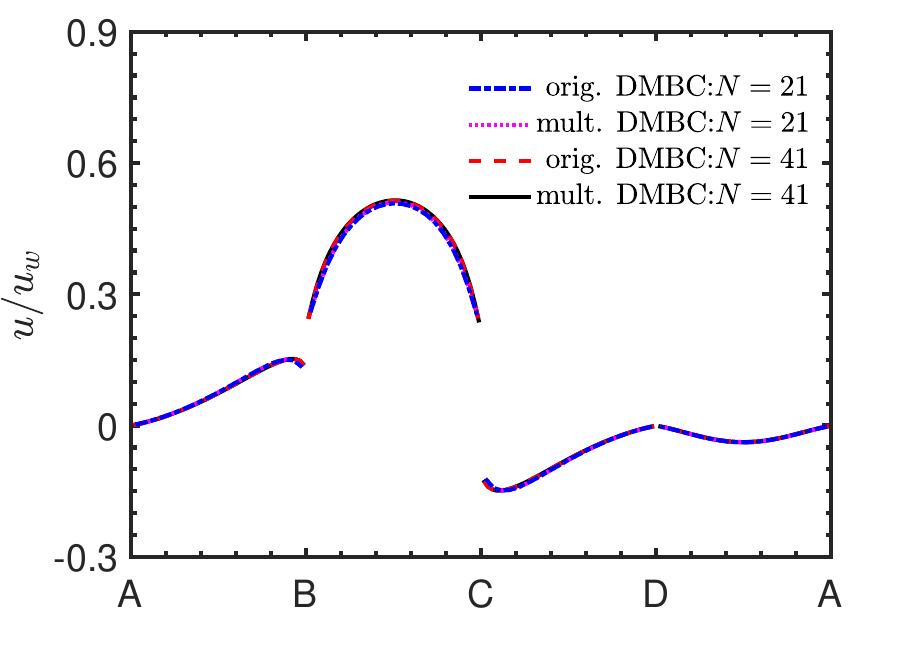}\label{FIG_Kn0075_uslip_TMAC05}}
{\includegraphics[width=0.32\textwidth]{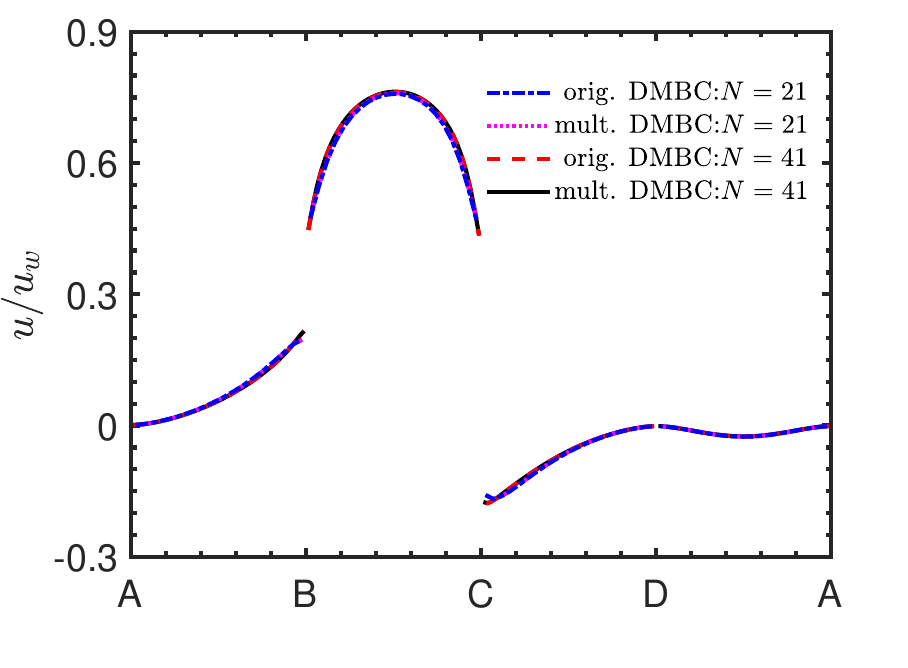}\label{FIG_Kn0075_uslip_TMAC1}}\\
{\includegraphics[width=0.32\textwidth]{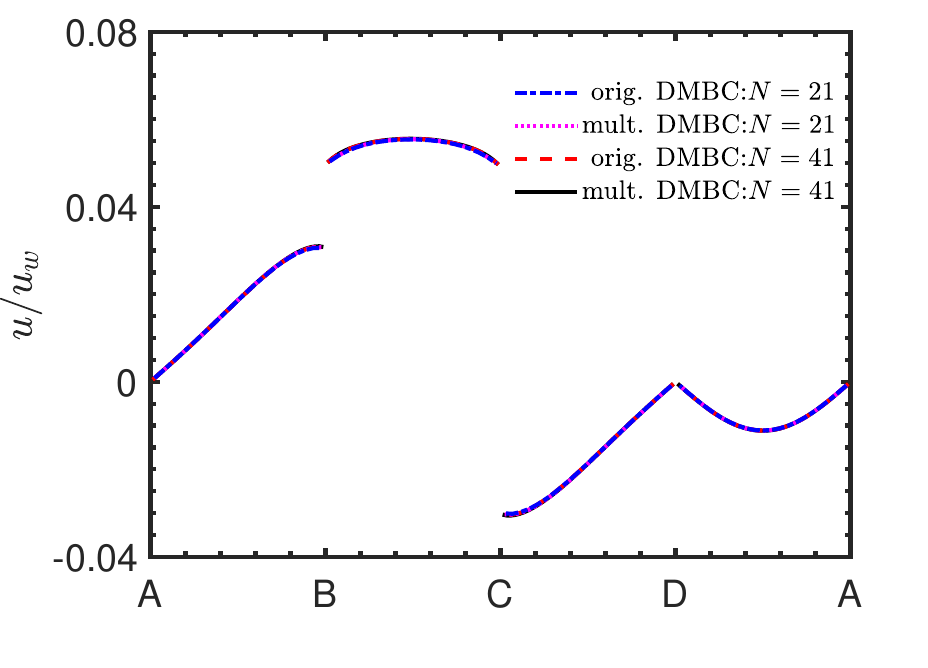}\label{FIG_Kn1_uslip_TMAC01}}
{\includegraphics[width=0.32\textwidth]{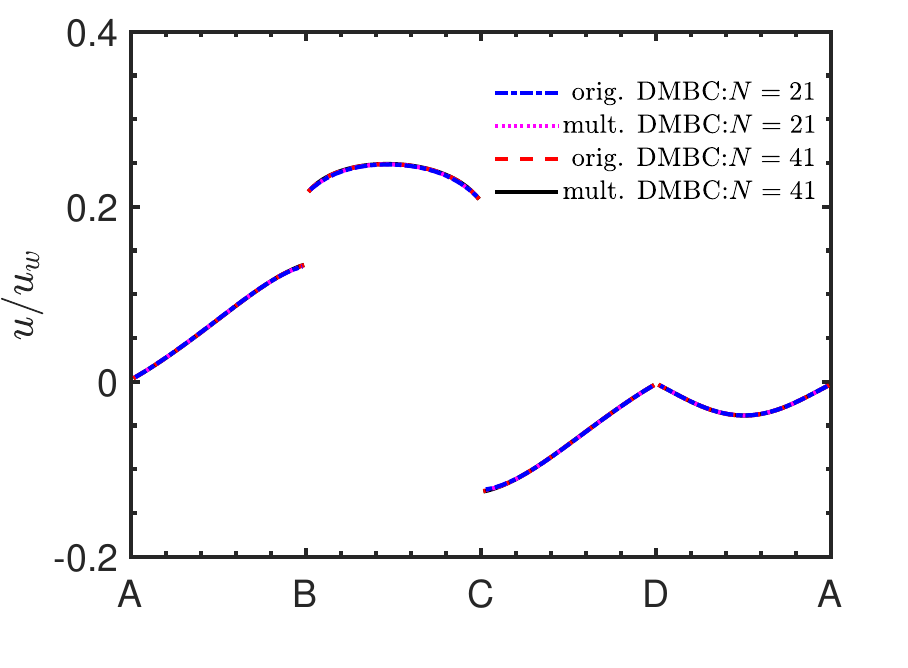}\label{FIG_Kn1_uslip_TMAC05}}
{\includegraphics[width=0.32\textwidth]{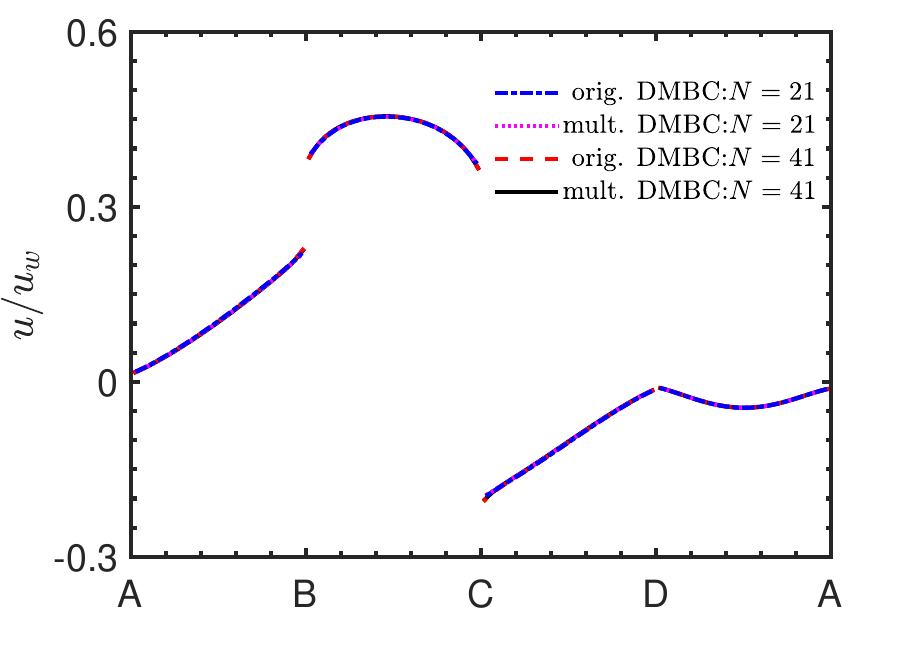}\label{FIG_Kn1_uslip_TMAC1}}\\
{\includegraphics[width=0.32\textwidth]{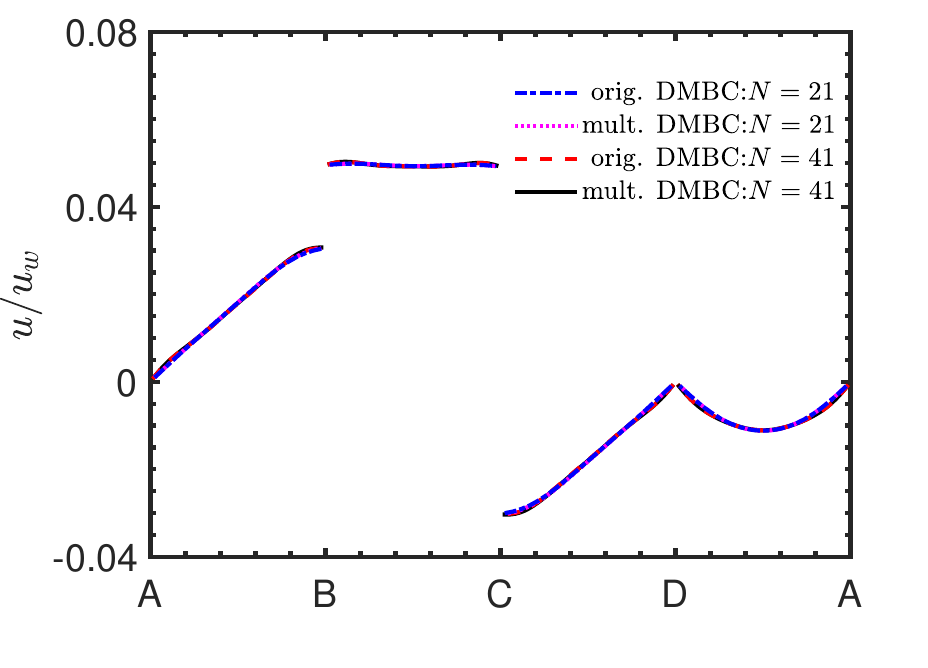}\label{FIG_Kn10_uslip_TMAC01}}
{\includegraphics[width=0.32\textwidth]{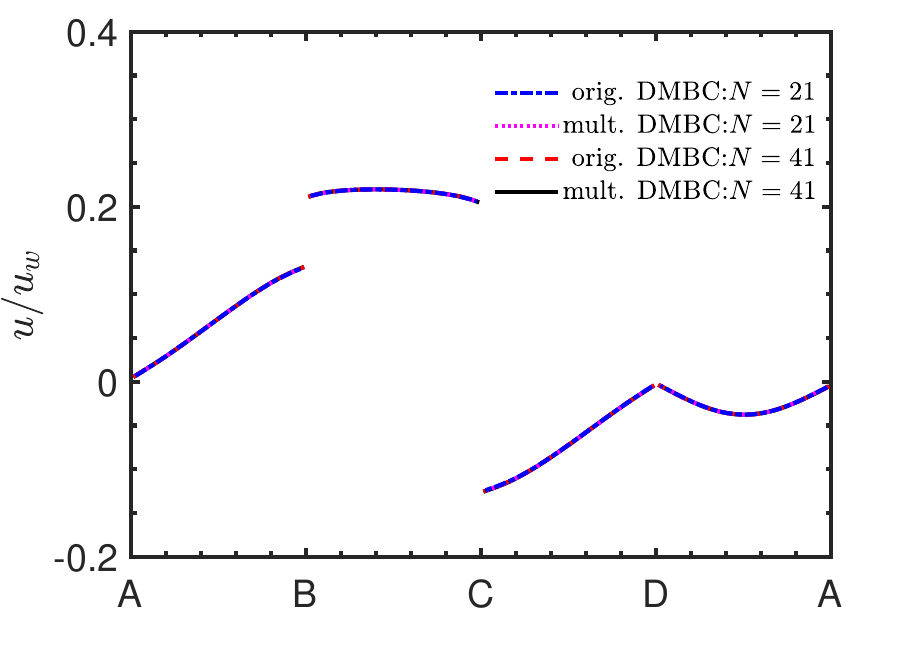}\label{FIG_Kn10_uslip_TMAC05}}
{\includegraphics[width=0.32\textwidth]{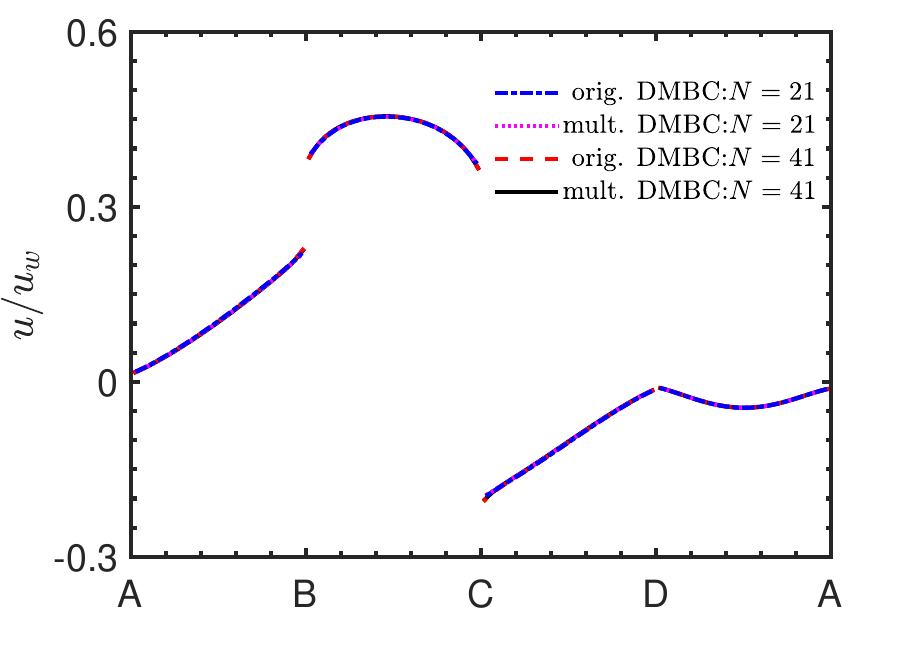}\label{FIG_Kn10_uslip_TMAC1}}
\caption{Tangential velocity profiles along the four walls of the cavity flow with different meshes and different accommodation coefficients ($\alpha$ from left to right: 0.1, 0.5 and
1.0) for different Kn: $\text{Kn} = 0.075$ (1st row), $\text{Kn} = 0.1$ (2nd row), $\text{Kn} = 10$ (3rd row). The benchmark data are from Ref~\cite{john2011effects}.}
\label{FIG:cavity uslip Kn}
\end{figure}

First, we consider the cases of $\text{Kn} = 0.075$, $1$, and $10$ for rarefied flow regime, corresponding to $\tau \approx 8.5 \times 10^{-2}$, $0.11$, and $1.1$, respectively. For validation, the velocity profiles along the vertical centerline and horizontal centerline of the cavity with different accommodation coefficients ($\alpha = 0.1$, $0.5$, and $1.0$) are presented in Fig.~\ref{FIG:cavity velocity Kn}, where the benchmark data are obtained from DSMC with fully diffuse-reflection conditions ($\alpha = 1$)\cite{john2011effects}. 
The results are almost identical on different mesh sizes and agree very well with the benchmark data. It is found that these two methods can give accurate results with just 21 meshes. In addition, the tangential velocity profiles along the four walls characterized by velocity slip are presented in Fig.~\ref{FIG:cavity uslip Kn}. The tangential velocity profiles from both schemes are indistinguishable along the wall. Similarly, the results of the stream function and primary vortex center, obtained with different mesh sizes and accommodation coefficients for both DMBCs, are consistent but not shown here. Therefore, it is confirmed that the both DMBCs are nearly identical in the rarefied flow regime. This is because the resulting spurious velocity slip is negligible as $\tau \gg h$. For the case of $N=21$, the ratio of $\tau$ to $h$ are 205.1, 2641.1, and 26411.5 for Knudsen numbers of 0.075, 1, and 10, respectively.

\begin{figure}[htbp]
\centering
{\includegraphics[width=0.45\textwidth]{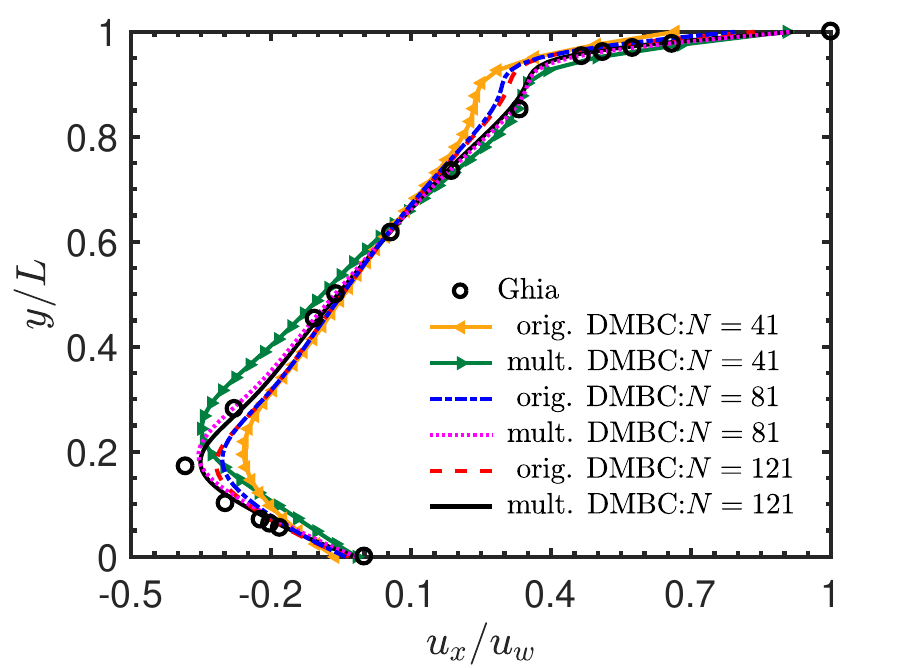}\label{FIG_Re1000_ux_TMAC01}}
{\includegraphics[width=0.45\textwidth]{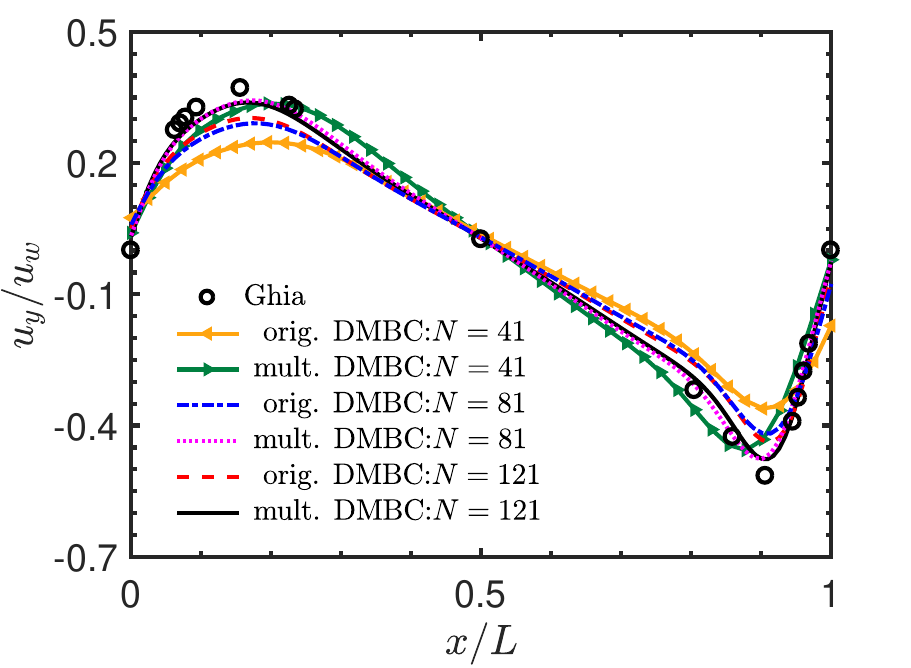}\label{FIG_Re1000_uy_TMAC01}}\\
{\includegraphics[width=0.45\textwidth]{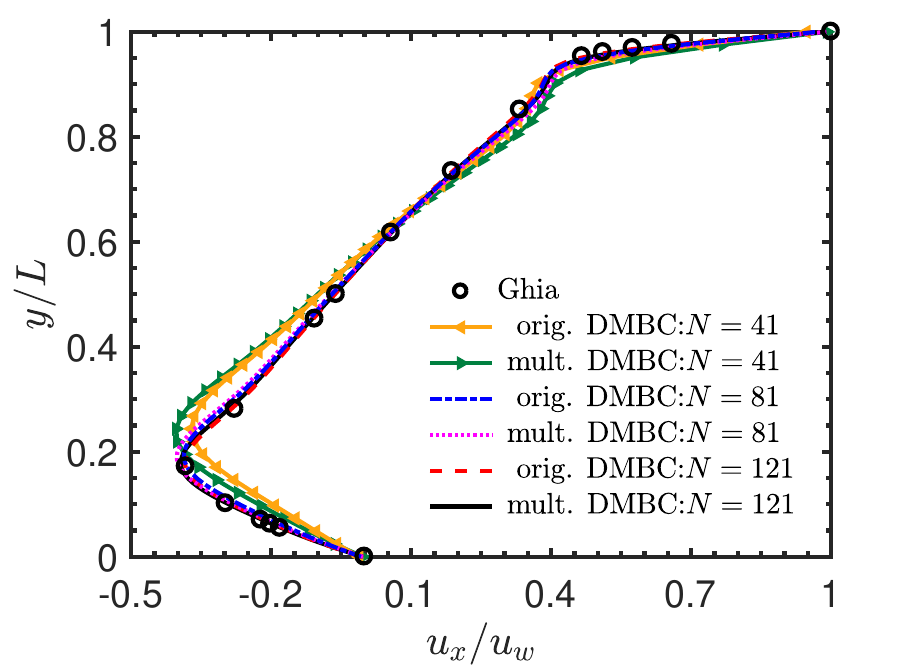}\label{FIG_Re1000_ux_TMAC05}}
{\includegraphics[width=0.45\textwidth]{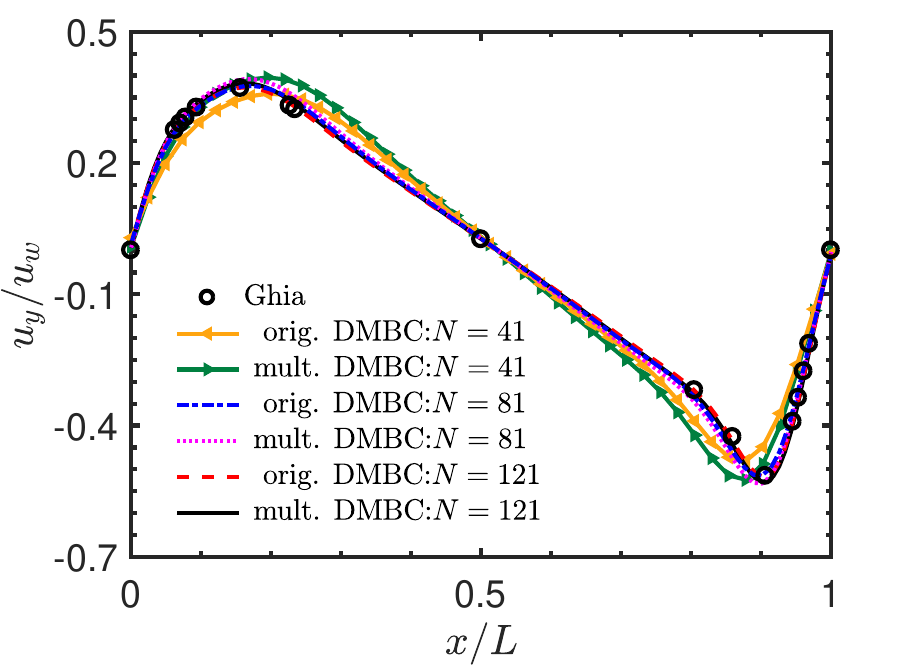}\label{FIG_Re1000_uy_TMAC05}}\\
{\includegraphics[width=0.45\textwidth]{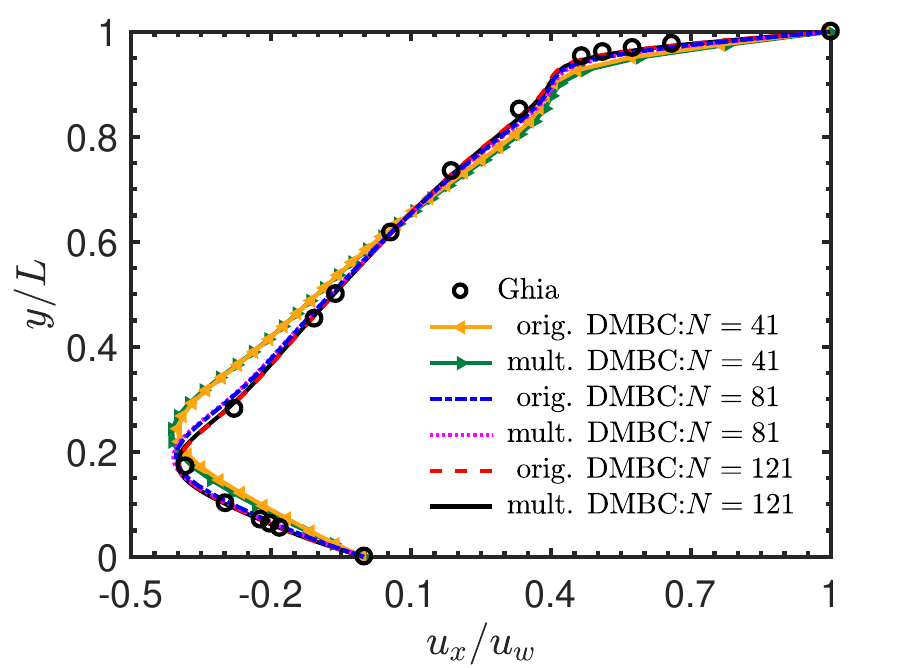}\label{FIG_Re1000_ux_TMAC1}}
{\includegraphics[width=0.45\textwidth]{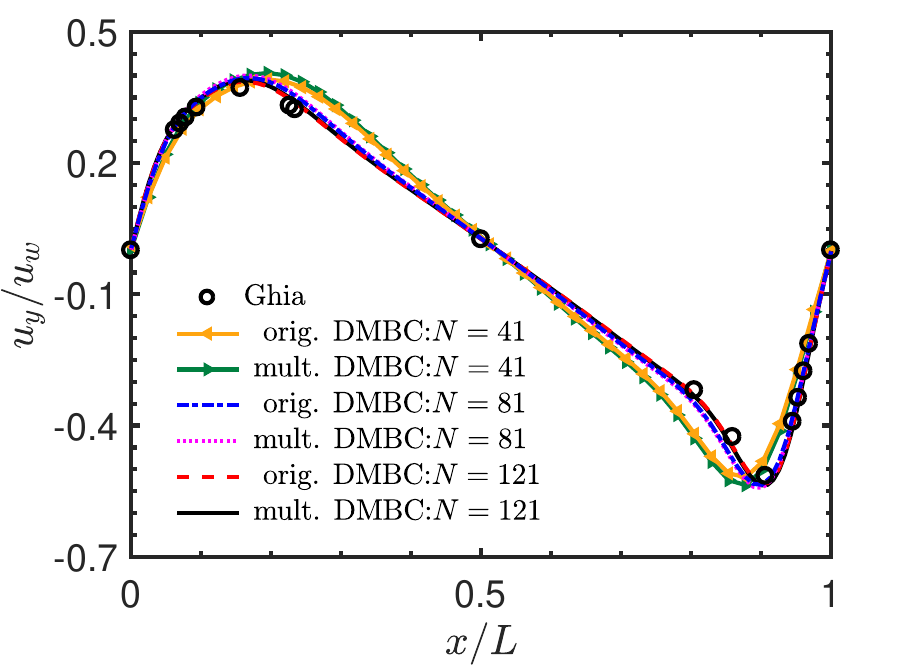}\label{FIG_Re1000_uy_TMAC1}}
\caption{Velocity profiles along the vertical centerline (1st column) and horizontal centerline (2nd column) of the cavity at $\text{Re}=1000$ with different meshes for different accommodation coefficients: $\alpha = 0.1$ (1st row), $\alpha = 0.5$ (2nd row), $\alpha = 1.0$ (3rd row). The benchmark data are from Ref~\cite{ghia1982high}.}
\label{FIG:cavity velocity Re1000}
\end{figure}

\begin{figure}[htbp]
\centering
\subfigure[]{\includegraphics[width=0.32\textwidth]{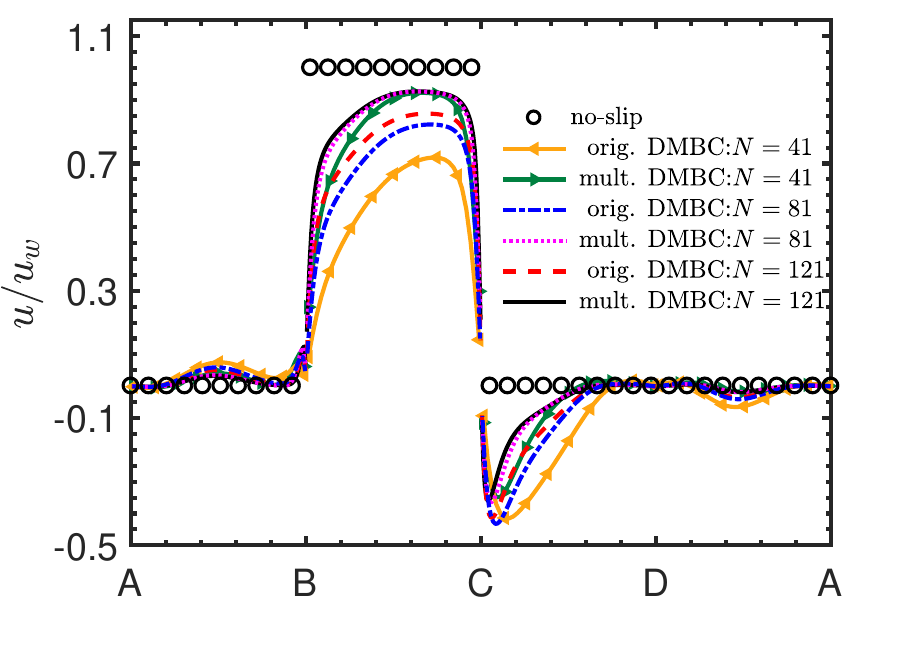}\label{FIG_Re1000_uslip_TMAC01}}
\subfigure[]{\includegraphics[width=0.32\textwidth]{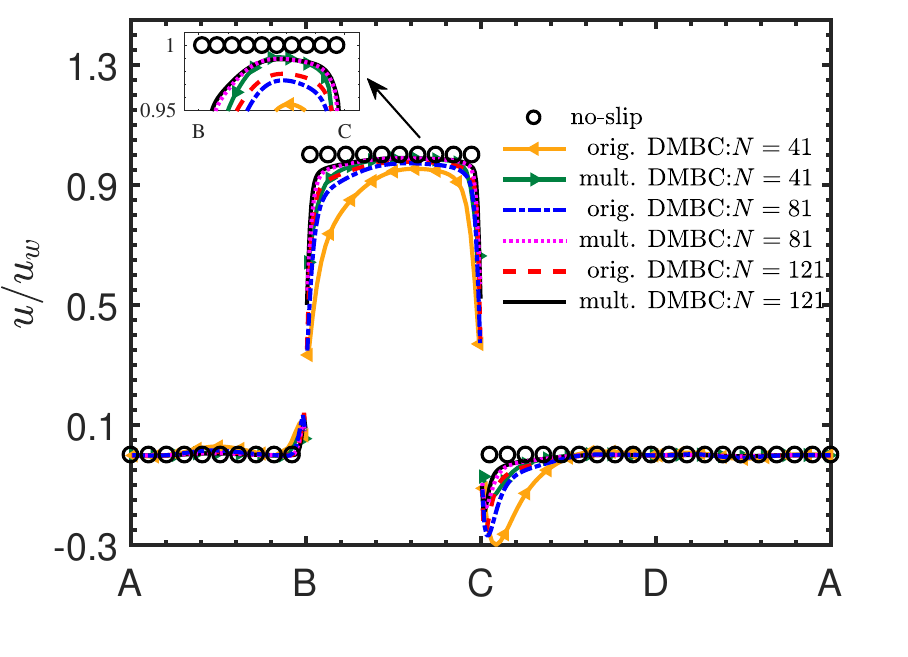}\label{FIG_Re1000_uslip_TMAC05}}
\subfigure[]{\includegraphics[width=0.32\textwidth]{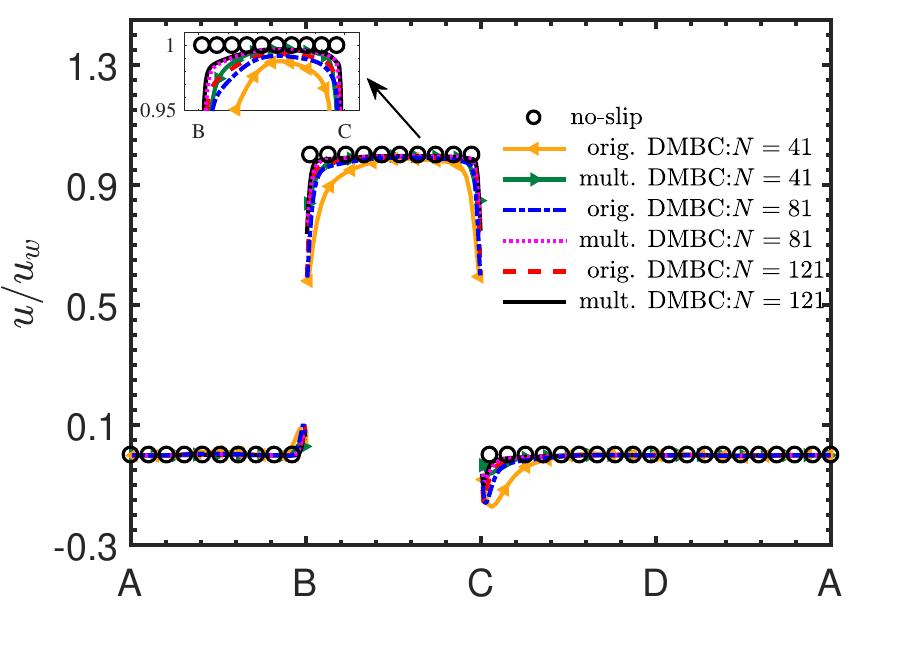}\label{FIG_Re1000_uslip_TMAC1}}
\caption{Tangential velocity profiles along the four walls of the cavity flow at $\text{Re}=1000$ with different meshes for different accommodation coefficients: (a) $\alpha = 0.1$, (b) $\alpha = 0.5$, (c) $\alpha = 1.0$.}
\label{FIG:cavity uslip Re1000}
\end{figure}

Then, we consider the case with the Reynolds number $\text{Re}=\rho_0 u_w L/\mu=1000$. In this case, the corresponding Knudsen number is about $5.2\times 10^{-5}$ and $\tau$ is about $ 3.2 \times 10^{-4}$.  
For validation, the velocity profiles along the vertical centerline and horizontal centerline of the cavity with different accommodation coefficients ($\alpha = 0.1$, $0.5$, and $1.0$) are presented in Fig.~\ref{FIG:cavity velocity Re1000}, where the benchmark NS solutions with no-slip BC are taken from Ghia~\cite{ghia1982high}. In order to compare the accuracy between the original and multiscale DMBCs, the results on different mesh resolutions are compared. As mesh resolution increased, the values of $\tau/h$ are approximately $0.12$, $0.24$, and $0.36$, respectively. As we can see in Fig.~\ref{FIG:cavity velocity Re1000}, the results of the multiscale DMBC with a mesh of $N=81$ show good agreement with benchmark data for all $\alpha$. However, when $\alpha =0.1$, the velocity profiles obtained using the original DMBC deviate significantly from the benchmark solution for all mesh sizes. As previously analyzed in Sec.~\ref{Continuum limit} theoretically, the original DMBC induces a spurious velocity slip proportional to $(2-\alpha)/\alpha$ and time step (or cell size). Consequently, as $\alpha$ increases, the results of the original DMBC with meshes of $N=81$ and $N=121$ agree well with the benchmark data at $\alpha = 0.5$ and $\alpha = 1.0$. As expected, a distinct difference in velocity slip at the wall between the two DMBCs is evident in Fig.~\ref{FIG:cavity uslip Re1000}. The multiscale DMBC with coarser meshes achieves the no-slip condition accurately. On the contrary, the original DMBC gives a spurious velocity slip at the lid, which decreases as the mesh is refined. It is worth noting that the top two corner points (point B and point C) appear singularly caused by a discontinuity between the stationary and moving walls. Obviously, the results of the multiscale DMBC at the corner points are better than those of the original one. 

\begin{figure}[htbp]
\centering
{\includegraphics[width=0.3\textwidth]{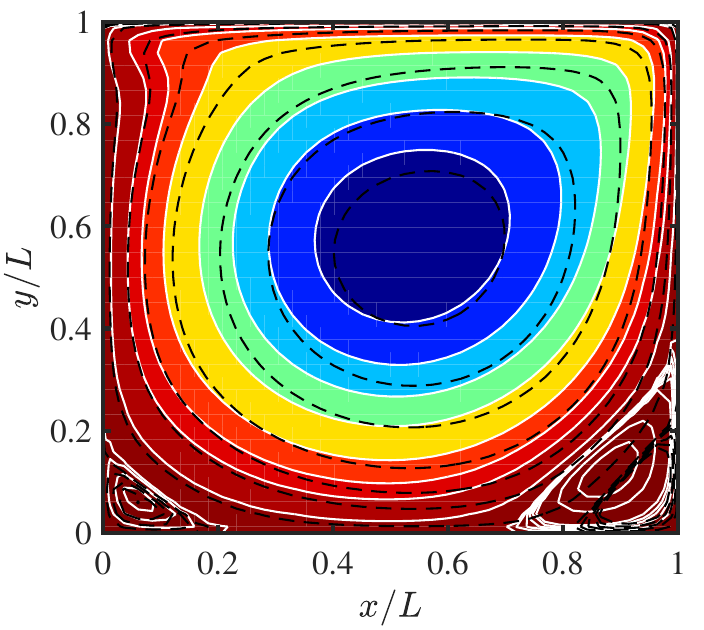}\label{FIG_Re1000_contour_psi_TMAC01N41}}
{\includegraphics[width=0.3\textwidth]{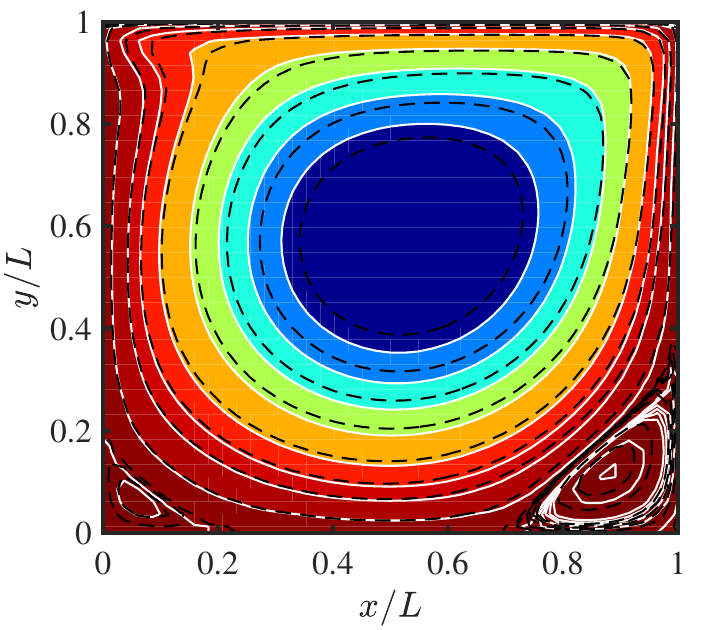}\label{FIG_Re1000_contour_psi_TMAC05N41}}
{\includegraphics[width=0.3\textwidth]{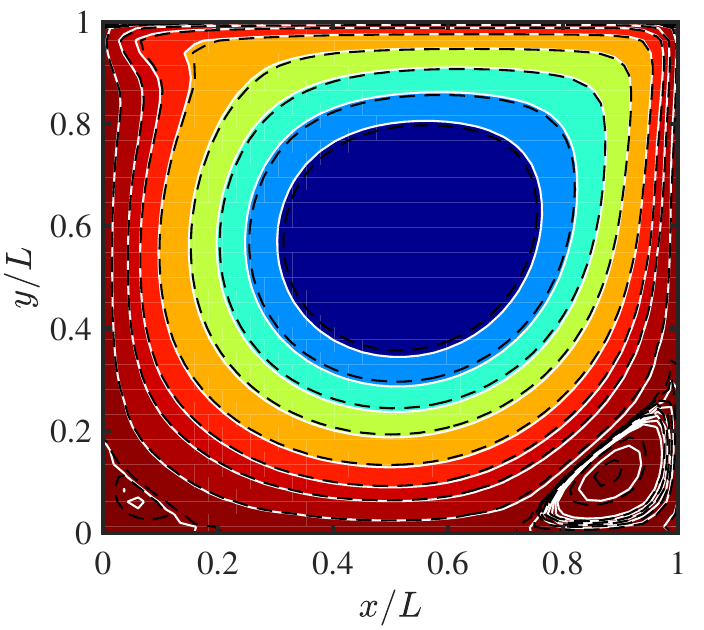}\label{FIG_Re1000_contour_psi_TMAC1N41}}
{\includegraphics[width=0.05\textwidth]{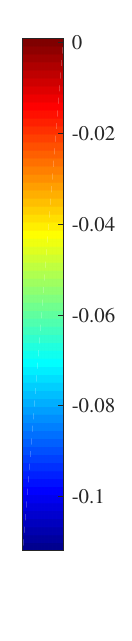}}
\\
{\includegraphics[width=0.3\textwidth]{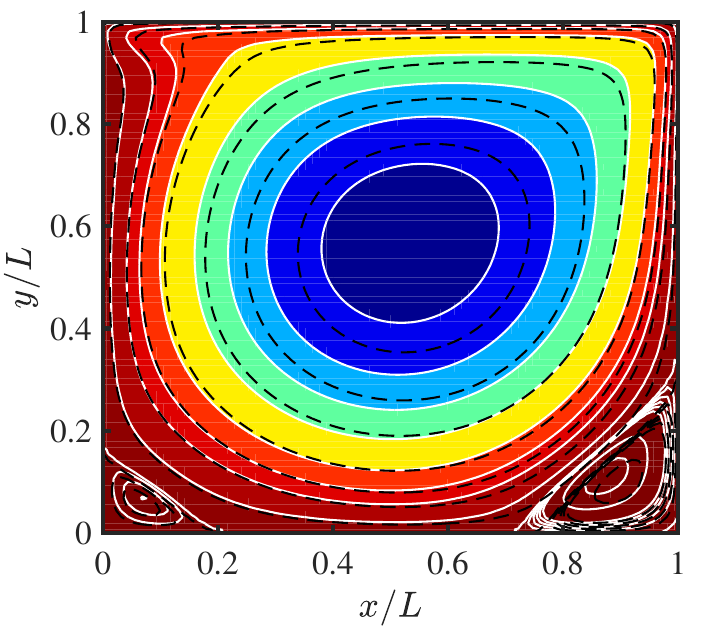}\label{FIG_Re1000_contour_psi_TMAC01N81}}
{\includegraphics[width=0.3\textwidth]{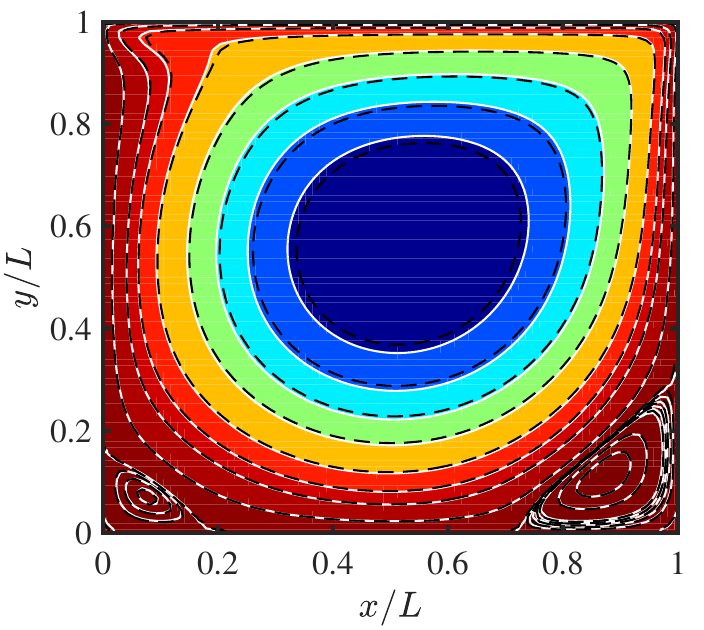}\label{FIG_Re1000_contour_psi_TMAC05N81}}
{\includegraphics[width=0.3\textwidth]{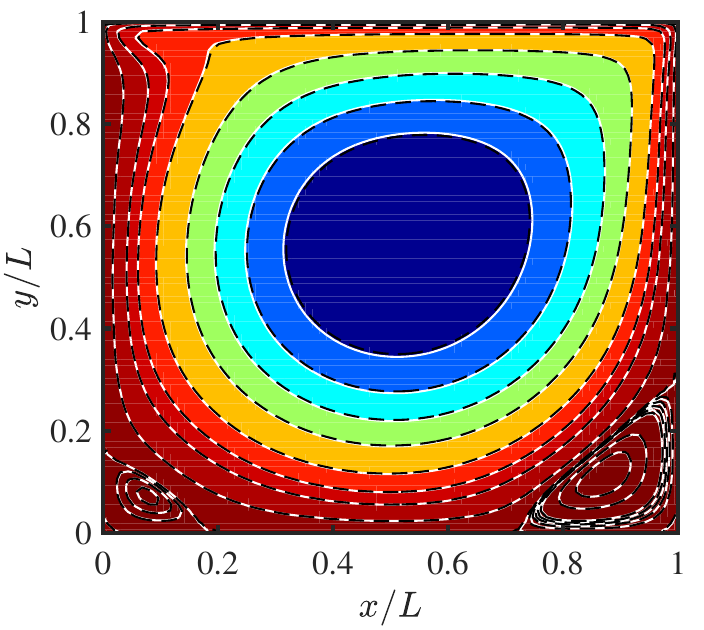}\label{FIG_Re1000_contour_psi_TMAC1N81}}
{\includegraphics[width=0.05\textwidth]{./Figures/legend.pdf}}
\\
{\includegraphics[width=0.3\textwidth]{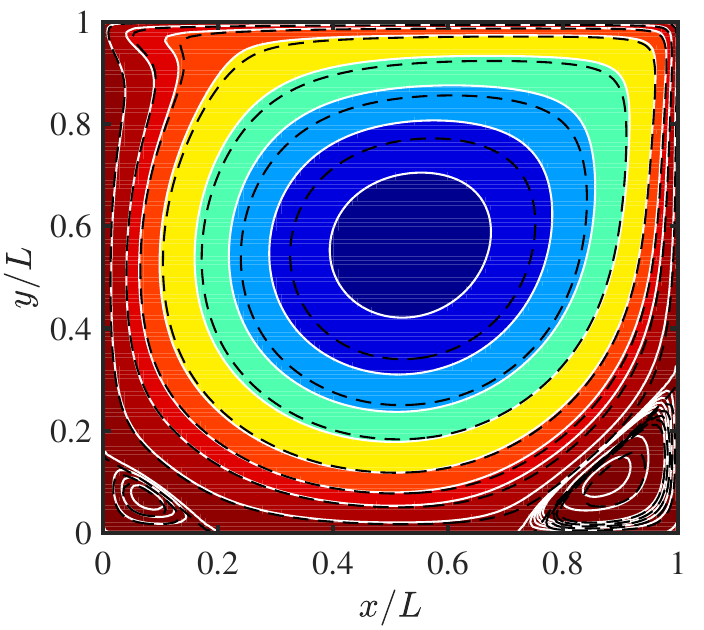}\label{FIG_Re1000_contour_psi_TMAC01N121}}
{\includegraphics[width=0.3\textwidth]{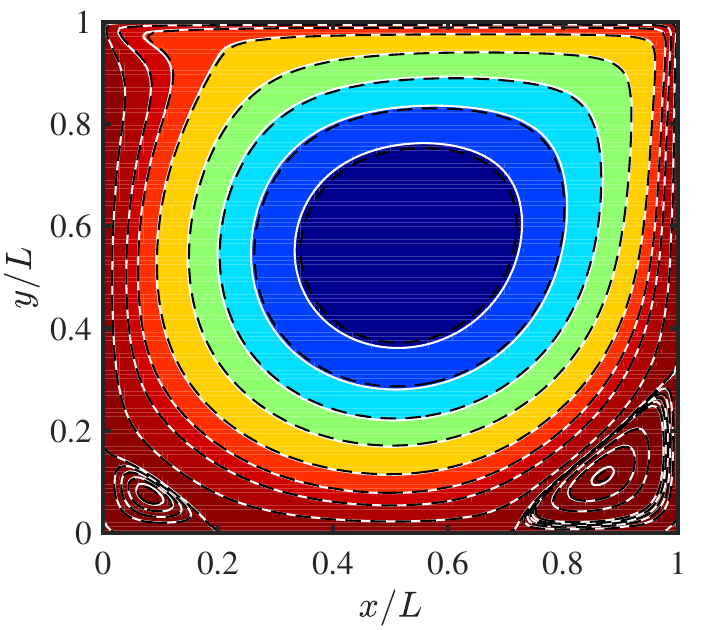}\label{FIG_Re1000_contour_psi_TMAC05N121}}
{\includegraphics[width=0.3\textwidth]{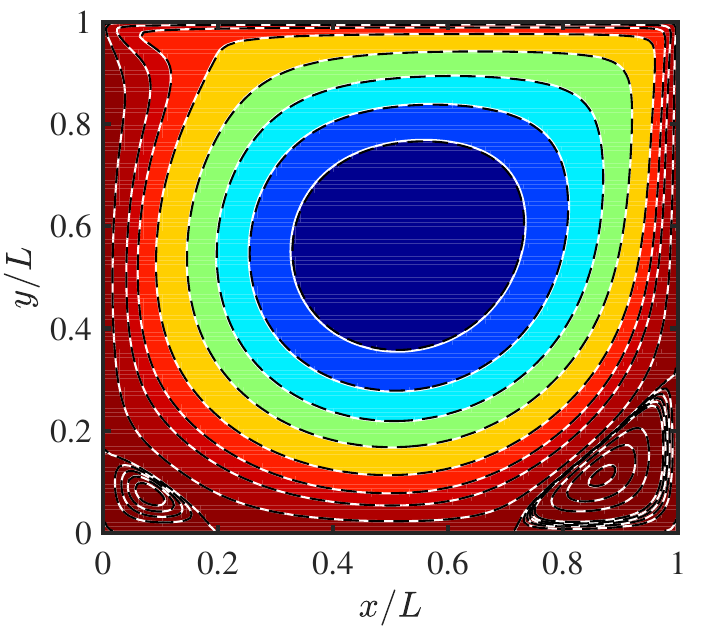}\label{FIG_Re1000_contour_psi_TMAC1N121}}
{\includegraphics[width=0.05\textwidth]{./Figures/legend.pdf}}
\caption{Stream function at $\text{Re} = 1000$ with different meshes (mesh sizes from top to bottom: 41, 81 and 121) for different accommodation coefficient: $\alpha = 0.1$ (1st column), $\alpha = 0.5$ (2nd  column), and $\alpha = 0.1$ (3rd column). Original DMBC: black dashed
lines; Multiscale DMBC: background with white solid lines.}
\label{FIG:cavity contour Re1000}
\end{figure}

Furthermore, we present the contours of the stream function at $\text{Re}=1000$  with different mesh sizes and accommodation coefficients for both DMBCs in Fig.~\ref{FIG:cavity contour Re1000}. As expected, the overall flow patterns reveal that both DMBCs yield similar results on the finest meshes with large values of $\alpha$, but exhibit clear differences on coarser meshes with smaller values of $\alpha$. Specifically, it can be observed that in the case of $N=40$ or $\alpha =0.1$, the results display significant differences, particularly in the vortices in the vicinity of the bottom right corner, where the vortices predicated the original DMBC are notably smaller. 

To further compare the flow fields quantitatively, we tabulate the positions of the primary and secondary lower vortex centers in Table~\ref{Table:Primary vortex}-\ref{Table:lower right vortex} together with the intensity of the vorticity $\omega$ at the vortex centers. It can be seen that the vortex center position and vorticity $\omega$  predicated by the multiscale DMBC agree well with the benchmark. The results obtained from the multiscale DMBC are also less sensitive to both mesh sizes and accommodation coefficients. On the other hand, the original DMBC gives quite similar but less accurate results on the finest meshes. However, The predicted errors in the vortex center position and vorticity increase rapidly, especially in the region of the two lower secondary vortices. This occurs because the original DMBC induces spurious velocity slip that enlarges the Knudsen layer, which should ideally be much thinner than the cell size in continuum flow. The enlarged Knudsen layer consequently smears the eddies near the boundary. 

The above observations confirm that both DMBCs yield similar results with sufficiently fine meshes, whereas on coarser meshes, the multiscale DMBC clearly demonstrates better accuracy. Additionally, the results of the multiscale DMBC are less sensitive to mesh resolution and accommodation coefficient for continuum flows. These results are consistent with the presented in Sec.~\ref{Continuum limit}.

\begin{table}[]
\caption{Primary vortex center position ($x$, $y$) and intensity of vorticity $\omega$ at the vortex center for the cavity flow at $\text{Re} = 1000$.}
\label{Table:Primary vortex}
\resizebox{\textwidth}{35mm}{
\begin{tabular}{lllllllllllll}
\hline
            & \multicolumn{1}{c}{$N$} & \multicolumn{1}{l}{($x$,} & \multicolumn{1}{r}{$y$)} & \multicolumn{1}{c}{$|\omega|$} &  & \multicolumn{1}{l}{($x$,} & \multicolumn{1}{r}{$y$)} & \multicolumn{1}{c}{$|\omega|$} &  & \multicolumn{1}{l}{($x$,} & \multicolumn{1}{r}{$y$)} & \multicolumn{1}{c}{$|\omega|$} \\ \cline{1-5} \cline{7-9} \cline{11-13} 
Ghia~\cite{ghia1982high} & 129 & (0.5313, & 0.5625) &2.049 &  & (0.5313, & 0.5625) &2.049 &  & (0.5313, & 0.5625) &2.049                                            \\ \cline{1-1} \cline{3-5} \cline{7-9} \cline{11-13} 
            & \multicolumn{1}{c}{}  & \multicolumn{3}{c}{$\alpha=0.1$}         &  & \multicolumn{3}{c}{$\alpha=0.5$} &  & \multicolumn{3}{c}{$\alpha=1.0$} \\ \cline{1-1} \cline{3-5} \cline{7-9} \cline{11-13} 
orig. DMBC  
            & \multicolumn{1}{c}{41}  & \multicolumn{1}{l}{(0.5490,} &\multicolumn{1}{r}{0.5605)}   & \multicolumn{1}{c}{1.635} & & \multicolumn{1}{l}{(0.5345,} &\multicolumn{1}{r}{0.5875)}   & \multicolumn{1}{c}{2.455}&  & \multicolumn{1}{l}{(0.5320,} &\multicolumn{1}{r}{0.5855)}   & \multicolumn{1}{c}{2.615}\\
            &\multicolumn{1}{c}{81} & \multicolumn{1}{l}{(0.5405,} &\multicolumn{1}{r}{0.5620)}   & \multicolumn{1}{c}{1.707} & & \multicolumn{1}{l}{(0.5315,} &\multicolumn{1}{r}{0.5700)}   & \multicolumn{1}{c}{2.164}&  & \multicolumn{1}{l}{(0.5300,} &\multicolumn{1}{r}{0.5695)}   & \multicolumn{1}{c}{2.245}\\
            &\multicolumn{1}{c}{121}  & \multicolumn{1}{l}{(0.5385,} &\multicolumn{1}{r}{0.5605)}   & \multicolumn{1}{c}{1.723} & & \multicolumn{1}{l}{(0.5315,} &\multicolumn{1}{r}{0.5665)}   & \multicolumn{1}{c}{2.081}&  & \multicolumn{1}{l}{(0.5305,} &\multicolumn{1}{r}{0.5665)}   & \multicolumn{1}{c}{2.141}\\
multi. DMBC 
            & \multicolumn{1}{c}{41}  & \multicolumn{1}{l}{(0.5350,} &\multicolumn{1}{r}{0.5730)}   & \multicolumn{1}{c}{2.087} & 
            &\multicolumn{1}{l}{(0.5335,} &\multicolumn{1}{r}{0.5875)}   
            &\multicolumn{1}{c}{2.114}&  & \multicolumn{1}{l}{(0.5315,} &\multicolumn{1}{r}{0.5724)}   & \multicolumn{1}{c}{2.155}\\
            &\multicolumn{1}{c}{81} & \multicolumn{1}{l}{(0.5313,} &\multicolumn{1}{r}{0.5679)}   & \multicolumn{1}{c}{1.955} & 
            &\multicolumn{1}{l}{(0.5315,} &\multicolumn{1}{r}{0.5679)}   
            &\multicolumn{1}{c}{2.064}&  & \multicolumn{1}{l}{(0.5311,} &\multicolumn{1}{r}{0.5675)}   & \multicolumn{1}{c}{2.072}\\
            &\multicolumn{1}{c}{121}  & \multicolumn{1}{l}{(0.5300,} &\multicolumn{1}{r}{0.5675)}   & \multicolumn{1}{c}{1.915} & 
            &\multicolumn{1}{l}{(0.5300,} &\multicolumn{1}{r}{0.5665)}   
            &\multicolumn{1}{c}{2.042}&  & \multicolumn{1}{l}{(0.5305,} &\multicolumn{1}{r}{0.5665)}   & \multicolumn{1}{c}{2.070}\\
            \hline
\end{tabular}}
\end{table}

\begin{table}[]
\caption{Positions of the lower left secondary vortex centers and intensity of vorticity $\omega$ at the vortex center for the cavity flow at $\text{Re} = 1000$.}
\label{Table:lower left vortex}
\resizebox{\textwidth}{35mm}{
\begin{tabular}{lllllllllllll}
\hline
            & \multicolumn{1}{c}{$N$} & \multicolumn{1}{l}{($x$,} & \multicolumn{1}{r}{$y$)} & \multicolumn{1}{c}{$|\omega|$} &  & \multicolumn{1}{l}{($x$,} & \multicolumn{1}{r}{$y$)} & \multicolumn{1}{c}{$|\omega|$} &  & \multicolumn{1}{l}{($x$,} & \multicolumn{1}{r}{$y$)} & \multicolumn{1}{c}{$|\omega|$} \\ \cline{1-5} \cline{7-9} \cline{11-13} 
Ghia~\cite{ghia1982high} & 129 & (0.0859, & 0.0781) &0.3618 &  & (0.0859, & 0.0781) &0.3618 &  & (0.0859, & 0.0781) &0.3618                                             \\ \cline{1-1} \cline{3-5} \cline{7-9} \cline{11-13} 
            & \multicolumn{1}{c}{}  & \multicolumn{3}{c}{$\alpha=0.1$}         &  & \multicolumn{3}{c}{$\alpha=0.5$} &  & \multicolumn{3}{c}{$\alpha=1.0$} \\ \cline{1-1} \cline{3-5} \cline{7-9} \cline{11-13} 
orig. DMBC  
            & \multicolumn{1}{c}{41}  & \multicolumn{1}{l}{(0.0620,} &\multicolumn{1}{r}{0.0515)}   & \multicolumn{1}{c}{0.2176} & & \multicolumn{1}{l}{(0.0625,} &\multicolumn{1}{r}{0.0665)}   & \multicolumn{1}{c}{0.1612}&  & \multicolumn{1}{l}{(0.0595,} &\multicolumn{1}{r}{0.0660)}   & \multicolumn{1}{c}{0.1308}\\
            &\multicolumn{1}{c}{81} & \multicolumn{1}{l}{(0.0650,} &\multicolumn{1}{r}{0.0595)}   & \multicolumn{1}{c}{0.1727} & & \multicolumn{1}{l}{(0.0760,} &\multicolumn{1}{r}{0.0710)}   & \multicolumn{1}{c}{0.2327}&  & \multicolumn{1}{l}{(0.0780,} &\multicolumn{1}{r}{0.076)}   & \multicolumn{1}{c}{0.2501}\\
            &\multicolumn{1}{c}{121}  & \multicolumn{1}{l}{(0.0701,} &\multicolumn{1}{r}{0.0635)}   & \multicolumn{1}{c}{0.2026} & & \multicolumn{1}{l}{(0.0795,} &\multicolumn{1}{r}{0.0740)}   & \multicolumn{1}{c}{0.2869}&  & \multicolumn{1}{l}{(0.0810,} &\multicolumn{1}{r}{0.0760)}   & \multicolumn{1}{c}{0.3097}\\
multi. DMBC 
            & \multicolumn{1}{c}{41}  & \multicolumn{1}{l}{(0.0825,} &\multicolumn{1}{r}{0.0775)}   & \multicolumn{1}{c}{0.3549} & 
            &\multicolumn{1}{l}{(0.0803,} &\multicolumn{1}{r}{0.0761)}   
            &\multicolumn{1}{c}{0.3314}&  & \multicolumn{1}{l}{(0.0796,} &\multicolumn{1}{r}{0.0721)}   & \multicolumn{1}{c}{0.3129}\\
            &\multicolumn{1}{c}{81} & \multicolumn{1}{l}{(0.0815,} &\multicolumn{1}{r}{0.0764)}   & \multicolumn{1}{c}{0.3461} & 
            &\multicolumn{1}{l}{(0.0804,} &\multicolumn{1}{r}{0.0750)}   
            &\multicolumn{1}{c}{0.3255}&  & \multicolumn{1}{l}{(0.0815,} &\multicolumn{1}{r}{0.0756)}   & \multicolumn{1}{c}{0.3301}\\
            &\multicolumn{1}{c}{121}  & \multicolumn{1}{l}{(0.0809,} &\multicolumn{1}{r}{0.0757)}   & \multicolumn{1}{c}{0.3375} & 
            &\multicolumn{1}{l}{(0.0824,} &\multicolumn{1}{r}{0.0765)}   
            &\multicolumn{1}{c}{0.3400}&  & \multicolumn{1}{l}{(0.0835,} &\multicolumn{1}{r}{0.0778)}   & \multicolumn{1}{c}{0.3429}\\
            \hline
\end{tabular}}
\end{table}

\begin{table}[]
\caption{Positions of the lower right secondary vortex centers  and intensity of vorticity $\omega$ at the vortex center for the cavity flow at $\text{Re} = 1000$.}
\label{Table:lower right vortex}
\resizebox{\textwidth}{35mm}{
\begin{tabular}{lllllllllllll}
\hline
            & \multicolumn{1}{c}{$N$} & \multicolumn{1}{l}{($x$,} & \multicolumn{1}{r}{$y$)} & \multicolumn{1}{c}{$|\omega|$} &  & \multicolumn{1}{l}{($x$,} & \multicolumn{1}{r}{$y$)} & \multicolumn{1}{c}{$|\omega|$} &  & \multicolumn{1}{l}{($x$,} & \multicolumn{1}{r}{$y$)} & \multicolumn{1}{c}{$|\omega|$} \\ \cline{1-5} \cline{7-9} \cline{11-13} 
Ghia~\cite{ghia1982high} & 129 & (0.8594, & 0.1094) &1.155 &  & (0.8594, & 0.1094) &1.155 & & (0.8594, & 0.1094) &1.155                                            \\ \cline{1-1} \cline{3-5} \cline{7-9} \cline{11-13} 
            & \multicolumn{1}{c}{}  & \multicolumn{3}{c}{$\alpha=0.1$}         &  & \multicolumn{3}{c}{$\alpha=0.5$} &  & \multicolumn{3}{c}{$\alpha=1.0$} \\ \cline{1-1} \cline{3-5} \cline{7-9} \cline{11-13} 
orig. DMBC  
            & \multicolumn{1}{c}{41}  & \multicolumn{1}{l}{(0.9335,} &\multicolumn{1}{r}{0.0825)}   & \multicolumn{1}{c}{0.4842} & & \multicolumn{1}{l}{(0.8755,} &\multicolumn{1}{r}{0.1205)}   & \multicolumn{1}{c}{0.8050}&  & \multicolumn{1}{l}{(0.8780,} &\multicolumn{1}{r}{0.1175)}   & \multicolumn{1}{c}{0.7962}\\
            &\multicolumn{1}{c}{81} & \multicolumn{1}{l}{(0.8970,} &\multicolumn{1}{r}{0.0984)}   & \multicolumn{1}{c}{0.6196} & & \multicolumn{1}{l}{(0.8715,} &\multicolumn{1}{r}{0.1115)}   & \multicolumn{1}{c}{0.9143}&  & \multicolumn{1}{l}{(0.8710,} &\multicolumn{1}{r}{0.1115)}   & \multicolumn{1}{c}{0.9553}\\
            &\multicolumn{1}{c}{121}  & \multicolumn{1}{l}{(0.8905,} &\multicolumn{1}{r}{0.1077)}   & \multicolumn{1}{c}{0.6972} & & \multicolumn{1}{l}{(0.8695,} &\multicolumn{1}{r}{0.1105)}   & \multicolumn{1}{c}{0.9809}&  & \multicolumn{1}{l}{(0.8685,} &\multicolumn{1}{r}{0.1110)}   & \multicolumn{1}{c}{1.021}\\
multi. DMBC 
            & \multicolumn{1}{c}{41}  & \multicolumn{1}{l}{(0.8745,} &\multicolumn{1}{r}{0.1207)}   & \multicolumn{1}{c}{0.9125} & 
            &\multicolumn{1}{l}{(0.8775,} &\multicolumn{1}{r}{0.1160)}   
            &\multicolumn{1}{c}{0.9637}&  & \multicolumn{1}{l}{(0.8815,} &\multicolumn{1}{r}{0.1150)}   & \multicolumn{1}{c}{1.0101}\\
            &\multicolumn{1}{c}{81} & \multicolumn{1}{l}{(0.8710,} &\multicolumn{1}{r}{0.1095)}   & \multicolumn{1}{c}{0.9538} & 
            &\multicolumn{1}{l}{(0.8690,} &\multicolumn{1}{r}{0.1115)}   
            &\multicolumn{1}{c}{1.0110}&  & \multicolumn{1}{l}{(0.8685,} &\multicolumn{1}{r}{0.1115)}   & \multicolumn{1}{c}{1.0350}\\
            &\multicolumn{1}{c}{121}  & \multicolumn{1}{l}{(0.8715,} &\multicolumn{1}{r}{0.1075)}   & \multicolumn{1}{c}{0.9712} & 
            &\multicolumn{1}{l}{(0.8665,} &\multicolumn{1}{r}{0.1105)}   
            &\multicolumn{1}{c}{1.0750}&  & \multicolumn{1}{l}{(0.8650,} &\multicolumn{1}{r}{0.1110)}   & \multicolumn{1}{c}{1.0923}\\
            \hline
\end{tabular}}
\end{table}

\section{Conclusion}\label{conclusion}
A multiscale discrete Maxwell boundary condition for the DUGKS is developed to predict the behavior of gas flow interacting with solid surfaces for all flow regimes. It can work properly across different regimes with a fixed discretization in both space and time, without limitations on the thickness of the Knudsen layer and relaxation time. Theoretical analysis and numerical tests (Couette flow, Fourier flow, and lid-driven cavity flow), are carried out to show the accuracy and capability of the proposed method.

It is found that discrete effects exist in the original discrete Maxwell boundary condition. Specifically, the original DMBC in DUGKS produce a spurious velocity slip and temperature jump, which are proportional to both the coefficient $(2-\alpha)/\alpha$ and time step (or cell size). Particularly, the slip velocity and temperature jump of the original DMBC first decrease and then remain constant. Based on the original DMBC, the multiscale DMBC is constructed by ensuring that the reflected original distribution function excludes collision effects.The numerical results demonstrate that the proposed multiscale DMBC works well for flows ranging from free molecular to continuum regimes. It is also found that the multiscale DMBC is less sensitive to mesh resolution and accommodation coefficient in the continuum regime. 

In conclusion, theoretical analysis and numerical results demonstrate that the DUGKS with the current multiscale DMBC is a promising method for simulating multiscale flows. It is worth noting that the aforementioned methods are not limited to DUGKS and Maxwell boundary condition, but are applicable to other kinetic methods and can be extended to other multiscale transport problems such as phonons transport and radiation transfer.

\section*{Acknowledgments}
This work was supported by the National Natural Science Foundation of China (Grant No. 12472290) and the Interdiciplinary Research Program of HUST (2023JCYJ002). 
Z. Xin gratefully acknowledges the support of the China Scholarship Council (No. 202406160103).
The computation is completed in the HPC Platform of Huazhong University of Science and Technology.

\newpage
\appendix

%\section{The Newton iterative method}
%The nonlinear system~\eqref{rho_w method II} and \eqref{eq: nonlinear W} can be rewritten as
%\begin{equation}
%    \boldsymbol G(\boldsymbol{\alpha})=0,
%\end{equation}
%where $\boldsymbol{\alpha}=\left(\rho_w, \rho^{n+1/2}, \boldsymbol u_t^{n+1/2}, T^{n+1/2}\right)^T$ is the unknown vector. In order to find solution for the above equation, the Newton iterative method is used by
%\begin{equation}
%    \boldsymbol \alpha^{k+1} = \boldsymbol \alpha^{k}-\left(\boldsymbol G'(\boldsymbol \alpha^{k})\right)^{-1}\boldsymbol G(\boldsymbol \alpha^{k}), 
%\end{equation}
%where 

\section{The multiscale DMBC based on Shakhov-BGK model in DUGKS} \label{appendix:MBC based on Shakhov-BGK model in DUGKS}
With Eq.~\eqref{eq:f x_w} and~\eqref{eq: f zero mass flux}, we can get
\begin{equation}
    \rho_w = -\sqrt{\frac{2\pi}{RT_w}}\left( \frac{2\tau}{2\tau+h}\left \langle \xi_n,\bar{f}_{in}^{n+1/2}(\boldsymbol \xi_{t}, \xi_{n}) \right \rangle _{<0} +\frac{h}{2\tau+h} \left \langle \xi_n, g^{n+1/2}(\boldsymbol x_w) \right  \rangle_{<0} \right),
\end{equation}
and
\begin{equation} 
\begin{aligned}
    \boldsymbol{W}^{n+1/2}(\boldsymbol{x}_w) =& \frac{2\tau}{2\tau+h} \left \langle \boldsymbol \psi ,\bar{f}_{in}^{n+1/2}(\boldsymbol \xi_{t}, \xi_{n})  \right \rangle_{<0} +  \frac{h}{2\tau+h} \left \langle \boldsymbol \psi, g^{n+1/2}(\boldsymbol x_w) \right  \rangle_{<0} 
    \\& + (1-\alpha)\left \langle \boldsymbol \psi ,f_{in}^{n+1/2}(\boldsymbol \xi_{t}, -\xi_{n})  \right \rangle_{>0} + \frac{\alpha}{2} \boldsymbol{I}_w .
\end{aligned}
\end{equation}

We can see the variables $\rho_w$ and $\boldsymbol{W}^{n+1/2}$ in the multiscale DMBC (Eqs.~\eqref{rho_w method II} and~\eqref{eq: nonlinear W}) depend on the half-range integral of $g$. When the equilibrium state $g = f^{S}$ is obtained by the Shakhov-BGK model, we can get,
\begin{equation}\label{A rho_w method II}
    \rho_w = \frac{h}{2\tau+h}\rho^{n+1/2}\sqrt{\frac{T^{n+1/2}}{T_w}} + \frac{4\tau}{2\tau+h}\frac{\left |\left \langle \xi_n,\bar{f}_{in}^{n+1/2}\right \rangle_{<0} \right| }{\sqrt{2RT_w/\pi}},
\end{equation}
and 
\begin{subequations} \label{A eq: nonlinear W}
\begin{align}
    \boldsymbol{W}^{n+1/2}(\boldsymbol{x}_w) =& \frac{4\tau}{4\tau+\alpha h} \left(\left \langle \boldsymbol \psi, \bar{f}_{in}^{n+1/2} \left( \boldsymbol \xi_t, \xi_n \right) \right \rangle_{<0}  + (1-\alpha) \left \langle \boldsymbol \psi, \bar{f}_{in}^{n+1/2} \left( \boldsymbol \xi_t, -\xi_n \right) \right \rangle_{>0}  \right)
    \notag \\ &+ \frac{\alpha (2\tau+h)}{4\tau+ \alpha h} \boldsymbol{I}_w +  \frac{2(2-\alpha)h}{4\tau+\alpha h} \left \langle \boldsymbol \psi, (f^{S}-f^{eq}) \right \rangle_{<0},\\
    q_n^{n+1/2} =& \frac{2\alpha\tau}{4\tau+(\alpha +2\text{Pr} - \alpha \text{Pr})h} \left \langle c_n c^2, \bar{f}_{in}^{n+1/2}\left( \boldsymbol \xi_t, \xi_n \right) \right \rangle_{<0}  \notag\\
    +& \frac{2\alpha h}{4\tau+(\alpha +2\text{Pr} - \alpha \text{Pr})h} \sqrt{\frac{2}{\pi}}\rho^{n+1/2}(RT^{n+1/2})^{3/2} \cdot \text{sign}\left( \left \langle c_n,1 \right \rangle_{<0} \right) \notag\\
    +&\frac{2\alpha(2\tau + h)}{4\tau+(\alpha +2\text{Pr} - \alpha \text{Pr})h} \sqrt{\frac{2}{\pi}}\rho_w(RT_w)^{3/2} \cdot \text{sign}\left( \left \langle c_n,1 \right \rangle_{>0} \right),\notag\\
    \boldsymbol{q}_t^{n+1/2} =&\frac{2\tau}{4\tau+(\alpha +2\text{Pr} - \alpha \text{Pr})h} \left \langle \boldsymbol c_t c^2, \bar{f}_{in}^{n+1/2}\left( \boldsymbol \xi_t, \xi_n \right) \right \rangle_{<0}  \notag\\
    +& \frac{2\tau(1-\alpha)}{4\tau+(\alpha +2\text{Pr} - \alpha \text{Pr})h} \left \langle \boldsymbol c_t c^2, \bar{f}_{in}^{n+1/2}\left( \boldsymbol \xi_t, -\xi_n \right) \right \rangle_{>0}
\end{align}
\end{subequations}
with
\begin{subequations}
    \begin{align}
        \left \langle 1, f^S-f^{eq} \right \rangle_{<0}&= -\frac{2(1-\text{Pr})q_n^{n+1/2}}{5\sqrt{\pi}(2RT^{n+1/2})^{3/2}}\cdot \text{sign}\left( \left \langle c_n,1 \right \rangle_{<0} \right),\\
        \left \langle  \xi_n ,f^S-f^{eq} \right \rangle_{<0}&= 0,\\
        \left \langle  \boldsymbol \xi_t, f^S-f^{eq} \right \rangle_{<0}&= -\frac{2(1-\text{Pr})q_n^{n+1/2} }{5\sqrt{\pi}(2R T^{n+1/2})^{3/2}} \boldsymbol{u}_t^{n+1/2}\cdot \text{sign}\left( \left \langle c_n ,1\right \rangle_{<0} \right),\\
        \left \langle  \frac{1}{2}\xi^2, f^S-f^{eq} \right \rangle_{<0}&= -\frac{(1-\text{Pr})q_n^{n+1/2} }{5\sqrt{\pi}(2R T^{n+1/2})^{3/2}} \left(\left|\boldsymbol u_t^{n+1/2}\right|^2 - 4RT^{n+1/2}\right)\cdot \text{sign}\left( \left \langle c_n ,1\right \rangle_{<0} \right),
    \end{align}
\end{subequations}
where, $c_n=\xi_n - u_n$ and $\boldsymbol c_t=\boldsymbol \xi_t - \boldsymbol u_t$ represent the normal and tangential component of peculiar velocity $\boldsymbol{c}$, respectively. Notably, $c_n<0$ means that the gas molecules are incident into the wall, and $c_n>0$ means that the gas molecules are reflected into the wall. But the positive or negative sign of the integral for its half of the interval should be judged according to the actual situation. The above system of equations with respect to the variables $\rho_w$,  $\boldsymbol W^{n+1/2}(\boldsymbol x_w)$ and $\boldsymbol q^{n+1/2}(\boldsymbol x_w)$ is nonlinear, which can be solved using certain iteration methods (the nonlinear Gauss-Seidel method is employed). Then, the interface flux at $\boldsymbol x_w$ can be determined based on Eq.~\eqref{eq:f x_w}.

\newpage
\bibliographystyle{elsarticle-num} 
\bibliography{myReferences}

\begin{thebibliography}{10}
\expandafter\ifx\csname url\endcsname\relax
  \def\url#1{\texttt{#1}}\fi
\expandafter\ifx\csname urlprefix\endcsname\relax\def\urlprefix{URL }\fi
\expandafter\ifx\csname href\endcsname\relax
  \def\href#1#2{#2} \def\path#1{#1}\fi

\bibitem{blanchard1997aerodynamic}
R.~C. Blanchard, R.~G. Wilmoth, J.~N. Moss, Aerodynamic flight measurements and rarefied-flow simulations of mars entry vehicles, Journal of Spacecraft and Rockets 34~(5) (1997) 687--690.

\bibitem{schouler2020survey}
M.~Schouler, Y.~Pr{\'e}vereaud, L.~Mieussens, Survey of flight and numerical data of hypersonic rarefied flows encountered in earth orbit and atmospheric reentry, Progress in Aerospace Sciences 118 (2020) 100638.

\bibitem{wang2014natural}
Q.~Wang, X.~Chen, A.~N. Jha, H.~Rogers, Natural gas from shale formation--the evolution, evidences and challenges of shale gas revolution in united states, Renewable and Sustainable Energy Reviews 30 (2014) 1--28.

\bibitem{moghaddam2016slip}
R.~N. Moghaddam, M.~Jamiolahmady, Slip flow in porous media, Fuel 173 (2016) 298--310.

\bibitem{karniadakis2006microflows}
G.~Karniadakis, A.~Beskok, N.~Aluru, Microflows and nanoflows: fundamentals and simulation, Vol.~29, Springer Science \& Business Media, 2006.

\bibitem{regmi2018micro}
B.~P. Regmi, M.~Agah, Micro gas chromatography: an overview of critical components and their integration, Analytical chemistry 90~(22) (2018) 13133--13150.

\bibitem{guo2014generalized}
Z.~Guo, J.~Qin, C.~Zheng, Generalized second-order slip boundary condition for nonequilibrium gas flows, Physical Review E 89~(1) (2014) 013021.

\bibitem{zeng2023second}
S.~Zeng, W.~Zhao, Z.~Jiang, W.~Chen, A second-order slip/jump boundary condition modified by nonlinear rayleigh--onsager dissipation factor, Physics of Fluids 35~(4) (2023).

\bibitem{chapman1990mathematical}
S.~Chapman, T.~G. Cowling, The mathematical theory of non-uniform gases: an account of the kinetic theory of viscosity, thermal conduction and diffusion in gases, Cambridge university press, 1990.

\bibitem{guo2013lattice}
Z.~Guo, C.~Shu, Lattice Boltzmann method and its application in engineering, Vol.~3, World Scientific, 2013.

\bibitem{bird1994molecular}
G.~A. Bird, Molecular gas dynamics and the direct simulation of gas flows, Oxford university press, 1994.

\bibitem{fan2001statistical}
J.~Fan, C.~Shen, Statistical simulation of low-speed rarefied gas flows, Journal of Computational Physics 167~(2) (2001) 393--412.

\bibitem{broadwell1964study}
J.~E. Broadwell, Study of rarefied shear flow by the discrete velocity method, Journal of Fluid Mechanics 19~(3) (1964) 401--414.

\bibitem{succi2002mesoscopic}
S.~Succi, Mesoscopic modeling of slip motion at fluid-solid interfaces with heterogeneous catalysis, Physical review letters 89~(6) (2002) 064502.

\bibitem{ansumali2002kinetic}
S.~Ansumali, I.~V. Karlin, Kinetic boundary conditions in the lattice boltzmann method, Physical Review E 66~(2) (2002) 026311.

\bibitem{guo2006physical}
Z.~Guo, T.~Zhao, Y.~Shi, Physical symmetry, spatial accuracy, and relaxation time of the lattice boltzmann equation for microgas flows, Journal of Applied physics 99~(7) (2006).

\bibitem{kim2008slip}
S.~H. Kim, H.~Pitsch, I.~D. Boyd, Slip velocity and knudsen layer in the lattice boltzmann method for microscale flows, Physical review E 77~(2) (2008) 026704.

\bibitem{liu2013lattice}
X.~Liu, Z.~Guo, A lattice boltzmann study of gas flows in a long micro-channel, Computers \& Mathematics with Applications 65~(2) (2013) 186--193.

\bibitem{su2017comparative}
W.~Su, S.~Lindsay, H.~Liu, L.~Wu, et~al., Comparative study of the discrete velocity and lattice boltzmann methods for rarefied gas flows through irregular channels, Physical Review E 96~(2) (2017) 023309.

\bibitem{mieussens2000discrete}
L.~Mieussens, Discrete velocity model and implicit scheme for the bgk equation of rarefied gas dynamics, Mathematical Models and Methods in Applied Sciences 10~(08) (2000) 1121--1149.

\bibitem{john2011effects}
B.~John, X.-J. Gu, D.~R. Emerson, Effects of incomplete surface accommodation on non-equilibrium heat transfer in cavity flow: A parallel dsmc study, Computers \& fluids 45~(1) (2011) 197--201.

\bibitem{meng2013assessment}
J.~Meng, L.~Wu, J.~M. Reese, Y.~Zhang, Assessment of the ellipsoidal-statistical bhatnagar--gross--krook model for force-driven poiseuille flows, Journal of Computational Physics 251 (2013) 383--395.

\bibitem{akhlaghi2023comprehensive}
H.~Akhlaghi, E.~Roohi, S.~Stefanov, A comprehensive review on micro-and nano-scale gas flow effects: Slip-jump phenomena, knudsen paradox, thermally-driven flows, and knudsen pumps, Physics Reports 997 (2023) 1--60.

\bibitem{jin1999efficient}
S.~Jin, Efficient asymptotic-preserving (ap) schemes for some multiscale kinetic equations, SIAM Journal on Scientific Computing 21~(2) (1999) 441--454.

\bibitem{filbet2010class}
F.~Filbet, S.~Jin, A class of asymptotic-preserving schemes for kinetic equations and related problems with stiff sources, Journal of Computational Physics 229~(20) (2010) 7625--7648.

\bibitem{dimarco2013asymptotic}
G.~Dimarco, L.~Pareschi, Asymptotic preserving implicit-explicit runge--kutta methods for nonlinear kinetic equations, SIAM Journal on Numerical Analysis 51~(2) (2013) 1064--1087.

\bibitem{guo2023unified}
Z.~Guo, J.~Li, K.~Xu, Unified preserving properties of kinetic schemes, Physical Review E 107~(2) (2023) 025301.

\bibitem{xu2010unified}
K.~Xu, J.-C. Huang, A unified gas-kinetic scheme for continuum and rarefied flows, Journal of Computational Physics 229~(20) (2010) 7747--7764.

\bibitem{huang2012unified}
J.-C. Huang, K.~Xu, P.~Yu, A unified gas-kinetic scheme for continuum and rarefied flows ii: multi-dimensional cases, Communications in Computational Physics 12~(3) (2012) 662--690.

\bibitem{Guo2013DiscreteUG}
Z.~Guo, K.~Xu, R.~Wang, Discrete unified gas kinetic scheme for all knudsen number flows: Low-speed isothermal case, Physical Review E 88~(3) (2013) 033305.

\bibitem{guo2015discrete}
Z.~Guo, R.~Wang, K.~Xu, Discrete unified gas kinetic scheme for all knudsen number flows. ii. thermal compressible case, Physical Review E 91~(3) (2015) 033313.

\bibitem{wang2015unified}
R.~Wang, Unified gas-kinetic scheme for the study of non-equilibrium flows, Ph.D. thesis, Hong Kong University of Science and Technology (2015).

\bibitem{yang2020analysis}
L.~Yang, Y.~Yu, L.~Yang, G.~Hou, Analysis and assessment of the no-slip and slip boundary conditions for the discrete unified gas kinetic scheme, Physical Review E 101~(2) (2020) 023312.

\bibitem{chen2022maxwell}
J.~Chen, S.~Liu, C.~Zhong, Y.~Yang, C.~Zhuo, Y.~Wang, D.~Jiang, The maxwell gas-surface interaction model for general discrete velocity framework in predictions of rarefied and multi-scale flows, arXiv preprint arXiv:2208.13992 (2022).

\bibitem{wang2015comparative}
P.~Wang, L.~Zhu, Z.~Guo, K.~Xu, A comparative study of lbe and dugks methods for nearly incompressible flows, Communications in Computational Physics 17~(3) (2015) 657--681.

\bibitem{zhang2018discrete}
Y.~Zhang, L.~Zhu, R.~Wang, Z.~Guo, Discrete unified gas kinetic scheme for all knudsen number flows. iii. binary gas mixtures of maxwell molecules, Physical Review E 97~(5) (2018) 053306.

\bibitem{zhu2019application}
L.~Zhu, Z.~Guo, Application of discrete unified gas kinetic scheme to thermally induced nonequilibrium flows, Computers \& Fluids 193 (2019) 103613.

\bibitem{tao2021application}
S.~Tao, L.~Wang, Y.~Ge, Q.~He, Application of half-way approach to discrete unified gas kinetic scheme for simulating pore-scale porous media flows, Computers \& Fluids 214 (2021) 104776.

\bibitem{gu2021computational}
Q.~Gu, M.-T. Ho, Y.~Zhang, Computational methods for pore-scale simulation of rarefied gas flow, Computers \& Fluids 222 (2021) 104932.

\bibitem{wang2022investigation}
Y.~Wang, S.~Liu, C.~Zhuo, C.~Zhong, Investigation of nonlinear squeeze-film damping involving rarefied gas effect in micro-electro-mechanical systems, Computers \& Mathematics with Applications 114 (2022) 188--209.

\bibitem{xin2023discrete}
Z.~Xin, Y.~Zhang, Z.~Guo, A discrete unified gas-kinetic scheme for multi-species rarefied flows, Advances in Aerodynamics 5~(1) (2023) 5.

\bibitem{wu2016discrete}
C.~Wu, B.~Shi, Z.~Chai, P.~Wang, Discrete unified gas kinetic scheme with a force term for incompressible fluid flows, Computers \& Mathematics with Applications 71~(12) (2016) 2608--2629.

\bibitem{maxwell1879vii}
J.~C. Maxwell, Vii. on stresses in rarified gases arising from inequalities of temperature, Philosophical Transactions of the royal society of London~(170) (1879) 231--256.

\bibitem{guo2007discrete}
Z.~Guo, B.~Shi, T.~Zhao, C.~Zheng, Discrete effects on boundary conditions for the lattice boltzmann equation in simulating microscale gas flows, Physical Review E 76~(5) (2007) 056704.

\bibitem{mieussens2013asymptotic}
L.~Mieussens, On the asymptotic preserving property of the unified gas kinetic scheme for the diffusion limit of linear kinetic models, Journal of Computational Physics 253 (2013) 138--156.

\bibitem{chen2015comparative}
S.~Chen, K.~Xu, A comparative study of an asymptotic preserving scheme and unified gas-kinetic scheme in continuum flow limit, Journal of Computational Physics 288 (2015) 52--65.

\bibitem{bhatnagar1954model}
P.~L. Bhatnagar, E.~P. Gross, M.~Krook, A model for collision processes in gases. i. small amplitude processes in charged and neutral one-component systems, Physical review 94~(3) (1954) 511.

\bibitem{holway1966new}
L.~H. Holway, New statistical models for kinetic theory: methods of construction, The physics of fluids 9~(9) (1966) 1658--1673.

\bibitem{shakhov1968generalization}
E.~Shakhov, Generalization of the krook kinetic relaxation equation, Fluid dynamics 3~(5) (1968) 95--96.

\bibitem{tao2020ghost}
S.~Tao, Q.~He, B.~Chen, F.~G. Qin, Y.~Lin, A ghost-cell discrete unified gas kinetic scheme for thermal flows with heat flux at curved interface, International Journal of Heat and Mass Transfer 162 (2020) 120365.

\bibitem{kremer2010introduction}
G.~M. Kremer, An introduction to the Boltzmann equation and transport processes in gases, Springer Science \& Business Media, 2010.

\bibitem{shizgal1981gaussian}
B.~Shizgal, A gaussian quadrature procedure for use in the solution of the boltzmann equation and related problems, Journal of Computational Physics 41~(2) (1981) 309--328.

\bibitem{zhang2012review}
W.-M. Zhang, G.~Meng, X.~Wei, A review on slip models for gas microflows, Microfluidics and nanofluidics 13 (2012) 845--882.

\bibitem{sone1990numerical}
Y.~Sone, S.~Takata, T.~Ohwada, Numerical analysis of the plane couette flow of a rarefied gas on the basis of the linearized boltzmann equation for hard-sphere molecules, European Journal of Mechanics B Fluids 9~(3) (1990) 273--288.

\bibitem{loyalka1975some}
S.~Loyalka, N.~Petrellis, T.~Storvick, Some numerical results for the bgk model: Thermal creep and viscous slip problems with arbitrary accomodation at the surface, Physics of Fluids 18~(9) (1975) 1094--1099.

\bibitem{wu2013deterministic}
L.~Wu, C.~White, T.~J. Scanlon, J.~M. Reese, Y.~Zhang, Deterministic numerical solutions of the boltzmann equation using the fast spectral method, Journal of Computational Physics 250 (2013) 27--52.

\bibitem{1968Momentum}
S.~K. Loyalka, Momentum and temperature‐slip coefficients with arbitrary accommodation at the surface, The Journal of Chemical Physics 48~(12) (1968) 5432--5436.

\bibitem{ghia1982high}
U.~Ghia, K.~N. Ghia, C.~Shin, High-re solutions for incompressible flow using the navier-stokes equations and a multigrid method, Journal of computational physics 48~(3) (1982) 387--411.

\end{thebibliography}





% \end{thebibliography}
\end{document}